\documentclass[a4paper,12pt]{article}
\usepackage{sola}
\usepackage{color}
\usepackage{todonotes}
\usepackage{subfigure}

\function{Gma}{\Gamma}

\defoperator{ii}{I}

\defoperator{QA}{Q^A}
\defoperator{QL}{Q^L}
\defoperator{QB}{Q^B}
\defoperator{QF}{Q^F}
\def\L{{\cal L}}

\defoperator{MA}{{\cal M}^\star}

\def\iig_#1[#2]{I_{#1}\br[{#2}]}
\defoperator{iigg}{I_g}
\def\iigf[#1]{\iigg[f,#1]}

\def\QAg_#1[#2]{Q^A_{#1}\br[{#2}]}
\defoperator{QAgg}{Q^A_g}
\def\QAgf[#1]{\QAgg[f,#1]}

\def\QLg_#1[#2]{Q^L_{#1}\br[{#2}]}
\defoperator{QLgg}{Q^L_g}
\def\QLgf[#1]{\QLgg[f,#1]}

\def\QBg_#1[#2]{Q^B_{#1}\br[{#2}]}
\defoperator{QBgg}{Q^B_g}
\def\QBgf[#1]{\QBgg[f,#1]}

\def\QFg_#1[#2]{Q^F_{#1}\br[{#2}]}
\defoperator{QFgg}{Q^F_g}
\def\QFgf[#1]{\QFgg[f,#1]}

\function{Lv}{\L[v]}

\def\sig_#1(#2){\sigma_{#1}\fpr(#2)}

\def\ewp(#1){\E^{\I \omega \pr({#1})}}
\def\ewg(#1){\E^{\I \omega g\pr(#1)}}

\def\sjacobiandd_#1^#2{\vc{J}_{#1}^{#2}}
\def\sjacobiand_#1^#2{\sjacobiandd_{#1}^{#2}(\xx)}

\def\sjacobian_#1{\sjacobiand_{#1}^d}

\def\fO(#1){\Oh\fpr({#1})}

\def\Leqpn^#1#2{\L[v]^{(#1)}(#2)=f^{(#1)}(#2)}
\def\Leqnpn^#1#2{\Leqpn^{#1}{x_#2}}

\def\xx{\vc x}

\def\Km_#1{{\cal K}_{#1}}
\defoperator{K}{{\cal K}_n}

\def\RHp{RH problem}

\def\Painleve{Painlev\'e}

\begin{document}

\setupproofs
\setupenvironments

\def\Tin{J_+^{-1}(z)}						
\def\Tout{J_-^{-1}(z)}						
\def\Tup{J_\uparrow^{-1}(x)}				
\def\Tdown{J_\downarrow^{-1}(x)}				
\def\Tupdown{J_\updownarrow^{-1}(x)}				

\refmodetrue

\newcommand{\sotodo}{\todo[color=green]}
\newcommand{\sotodoinline}{\todo[color=green,inline=true]}
\newcommand{\tttodo}{\todo[color=yellow]}
\newcommand{\tttodoinline}{\todo[color=yellow,inline=true]}
\newcommand{\goto}{\rightarrow}
\newcommand{\PII}{Painlev\'e~II}

\newenvironment{mat}{\left[\begin{array}{ccccccccccccccc}}{\end{array}\right]}
\newcommand\bcm{\begin{mat}}
\newcommand\ecm{\end{mat}}

\newcommand{\bigo}{\mathcal O}

\newenvironment{choices}{\left\{ \begin{array}{ll}}{\end{array}\right.}
\newcommand\when{&\text{if~}}
\newcommand\otherwise{&\text{otherwise}}

\def\Tr{{\rm Tr}\,}
\def\arctanh{{\rm arctanh}\,}
\def\dmu{{\dkf \mu}}

\def\KK{{\cal K}}

\authord={Sheehan Olver and Thomas Trogdon}

\titled={Numerical solution of Riemann--Hilbert problems: random matrix theory and orthogonal polynomials}

\maketitle

\Abstract
	In recent developments, a general approach for solving Riemann--Hilbert problems numerically has been developed.  We review this numerical framework, and apply it to the calculation of orthogonal polynomials on the real line.  Combining this numerical algorithm with an approach to compute Fredholm determinants, we are able to  calculate level densities and gap statistics for general finite-dimensional unitary ensembles.    We also include a description of how to compute the Hastings--McLeod solution of the homogeneous \Painleve\ II equation.
	
\Section{intro}  Introduction.

	We are concerned with calculating random matrix statistics for Hermitian invariant ensembles; i.e., $n \times n$ random matrices
	$$M = \sopmatrix{
		M_{11} & M_{12}^{\rm R} + \I M_{12}^{\rm I} &\cdots & M_{1n}^{\rm R} + \I M_{1n}^{\rm I} \cr
		 M_{12}^{\rm R} - \I M_{12}^{\rm I} & M_{22 }& \cdots & M_{2n}^{\rm R} + \I M_{2n}^{\rm I} \cr
		 \vdots & \ddots & \ddots & \vdots \cr
		 M_{1n}^{\rm R} - \I M_{1n}^{\rm I}  & \cdots &		 M_{(n-1)n}^{\rm R} - \I M_{(n-1)n}^{\rm I} & M_{nn}
	}$$
whose entries are distributed according to
	$${1 \over Z_n} \E^{-n \Tr V(M)} \dkf M,$$
where $Z_n$ is the normalization constant and
	$$\dkf M = \prod_{i = 1}^n \dkf M_{ii} \prod_{i < j} (\dkf M_{ij}^{\rm R} \dkf M_{ij}^{\rm I}).$$

	The eigenvalue statistics of invariant ensembles are expressible in terms of the kernel
	$$\KK_n(x,y) = -{\gamma_{n-1} \E^{-n/2(V(x) + V(y))}  \over 2\pi \I} {\pi_{n}(x) \pi_{n-1}(y) - \pi_{n-1}(x) \pi_{n}(y) \over x - y},\qquad\hbox{\cite{DeiftOrthogonalPolynomials}}$$
where $\pi_k$ are  the orthonormal polynomials $\pi_0,\pi_1,\ldots$ with respect to the weight
	$$\E^{- n V(x)} \dx,$$
and 
$$\gamma_{n-1} ={ 2 \pi \I} \br[ \int_{-1}^1 \pi_{n-1}(x) w(x) \dx]^{-1}$$
is a normalization constant.
%
%
Particular statistics include the {\it level density}
	$$\dmu_n = \KK_n(x,x) \dx,$$
describing the global distribution of eigenvalues, and the {\it gap statistic}
	$$\det(I - \KK_n|_{L^2[\Omega]}),$$
where $\det$ denotes a {\it Fredholm determinant}, describing the local distribution of eigenvalues; namely, the probability that no eigenvalue is inside the set $\Omega$.

Gap statistics for invariant ensembles follow two principles of universality.  For $x$ in   the {\it bulk} --- i.e., inside the support of the equilibrium measure --- the gap statistic of a properly scaled neighbourhood of $x$ will  approach the sinc kernel distribution:
	$$\det(I - {\cal S}|_{L^2(-s,s)})\qfor {\cal S} = {\sin(x-y) \over x - y}.$$
 This was proved rigorously in \cite{DeiftOrthogonalPolynomials} by expressing the orthogonal polynomials in terms of a {\it Riemann--Hilbert problem}, so that asymptotics of $\pi_n$ were determinable via nonlinear steepest descent.  Moreover, the {\it edge statistic} --- i.e., a properly scaled neighbourhood of $\infty$ --- generically approaches the Tracy--Widom distribution:
	$$\det(I - {\cal A}|_{L^2(s,\infty)})\qfor {\cal A} = {\Ai(x) \Ai'(y) - \Ai'(x) \Ai(y) \over x - y}.$$
Underlying these two universality laws are {\it \Painleve\ transcendents}; in the case of the Tracy--Widom distribution it is the Hastings--McLeod solution to \Painleve\ II \cite{HastingsMcLeod}, whereas the sine kernel distribution is expressible in terms of a solution to \Painleve\ V \cite{MehtaRM}.  See Section~\ref{appendix:PII} for a discussion of a numerical Riemann--Hilbert approach for computing the Hastings--McLeod solution of \Painleve\ II. 

	The statistics differ from universality laws for finite $n$ and are no longer expressible in terms of \Painleve\ transcendents.  Hence our aim is to calculate the finite-dimensional statistics to explore the manner in which the onset of universality depends on the potential $V$.  To accomplish this task, we will calculate the associated orthogonal polynomials numerically, also using their Riemann--Hilbert representation, via the framework of \cite{SOPainleveII,SORHFramework}.    By deforming the contours appropriately, we will achieve a numerical method that is uniformly accurate for large and small $n$, as shown in \cite{TrogdonSONNSD}.  
	
	 We will see in our numerical experiments that the onset of universality   depends strongly  on the magnitude of the equilibrium measure:  where eigenvalue density is small, finite $n$ statistics differ from universality behaviour greatly.  
	 

	We begin with a demonstration of the numerical calculated finite-dimensional random matrix statistics (\secref{RMT}).  Importantly, because we do not require the knowledge of local parametrices, our numerical approach continues to work for degenerate potentials, such as those that arise in the study of higher order Tracy--Widom distributions \cite{HigherOrderTracyWidom}.    We describe the manner in which orthogonal polynomials can be reduced to a Riemann--Hilbert problem that is suitable for numerics (\secref{RHP}).  We then review the numerical method for Riemann--Hilbert problems (\secref{nmRHP}), based on the deformations of \cite{DeiftOrthogonalPolynomials}.  This includes the result that the numerical approximation is {\it uniformly accurate} when the contours are  appropriately deformed (\secref{Uniform}), without the use of classical local parametrices.  


\Remark
An alternative to the approach advocated in this paper is to calculate the orthogonal polynomials directly for each $n$ via Gram--Schmidt and numerical quadrature.   For small $n$, this is likely to be more efficient.  However, it is well known to be prone to instability \cite{GautschiOP};  moreover, the calculation must be restarted for each $n$ as the weight $\E^{-n V}$ changes.  On the other hand, the RH approach has computational cost independent of $n$, making it more practical for investigating large $n$ behaviour.

\Section{RMT}  Random matrix theory.

Recall that
	$$\KK_n(x,y) = -{\gamma_{n-1} \E^{-n/2(V(x) + V(y))}  \over 2\pi \I} {\pi_{n}(x) \pi_{n-1}(y) - \pi_{n-1}(x) \pi_{n}(y) \over x - y}$$
and
	$$\KK_n(x,x) = -{\gamma_{n-1} \E^{-n V(x) }  \over 2\pi \I} (\pi_{n}'(x) \pi_{n-1}(x) - \pi_{n-1}'(x) \pi_{n}(x)).$$
In this section, we use the approach of numerically calculating $\pi_n$ and $\gamma_{n-1} \pi_{n-1}$ that we develop below to  compute the finite $n$ statistics.   What will be apparent in the numerical results is that the behaviour of local statistics is tied strongly to the global density of eigenvalues near the region; i.e., the magnitude of the level density.

\Figurethree{GaussEmpirical3}{GaussEmpirical10}{GaussEmpirical100}
	Calculated level densities for the GUE for $n = 3, 10$ and 100, compared to histograms.

For unitary invariant ensembles, the level density is the distribution of the counting measure.  This is precisely
	$$\dkf \mu_n = {\KK_n(x,x) \over n} \dx.$$
	In \figref{GaussEmpirical3}, we compare  the (numerically calculated) GUE (i.e., $V(x) = x^2$) level density  for $n = 3, 10$ and $ 100$ to a histogram, demonstrating the accuracy of the approximation.  (Because the polynomials involved are Hermite polynomials, we can also verify the accuracy directly.)  This shows the standard phenomena that the distribution exhibits $n$ ``bumps'' of increased density, corresponding to the positions of the finite charge  energy minimization equilibrium; i.e., the Fekete points.

\Figurethree{DegenerateEmpirical3}{DegenerateEmpirical10}{DegenerateEmpirical100}
	Calculated level density for  $V(x) = {x^2 \over 5} - {4 \over 15} x^3 + {x^4 \over 20} + {8 \over 5} x$ for $n = 3, 10$ and 100.  Dashed line is the equilibrium measure ($n = \infty$).

	In \figref{DegenerateEmpirical3}, we plot the finite $n$ level densities for the potential
	$$V(x) = {x^2 \over 5} - {4 \over 15} x^3 + {x^4 \over 20} + {8 \over 5} x,$$
which is an example of a potential whose equilibrium measure vanishes at an endpoint, and hence the edge statistics follow the higher order Tracy--Widom distribution \cite{HigherOrderTracyWidom}.     Interestingly, this change in edge statistic behaviour is not just present in the local statistics, but clearly visible in the decay of the tail of the global statistics.

We now turn our attention to local gap statistics, which are described by the \newterm{Fredholm determinant}
	$$\det(I - \KK_n|_{L^2[\Omega]}).$$
Using the method of Bornemann \cite{BornemannFredholm}, we can calculate the determinant, provided that the kernel itself can be evaluated.  Thus, we can successfully calculate finite gap statistics by using the RH approach to calculate $\pi_n$ and $\gamma_{n-1} \pi_{n-1}$.   In \figref{GaussGap50}, we plot the gap statistics versus a histogram for the GUE in the interval $(-s,s)$.

\Figuretwo{GaussGap50}{GaussGap100}
	The calculated probability that there are no eigenvalues in $(-s,s)$ for the GUE (plain) versus Monte Carlo simulation (dashed), for $n = 50$ (left) and $n = 100$ (right).

\Figuretwo{DegenerateGap1}{DegenerateGap15}
	The calculated probability that there are no eigenvalues in the scaled neighbourhood $x + {(-s,s) \over {\cal K}_n(x,x)}$ for $n = 50, 100, 200$ and 250 for $x = 1$ (left) and $x = 1.5$ (right), for the potential $V(x) = {x^2 \over 5} - {4 \over 15} x^3 + {x^4 \over 20} + {8 \over 5} x$.

To see universality in the bulk, we have to scale the interval with $n$; in particular, we need to look at the gap probability for 
	$$\Omega = x + {(-s,s) \over \KK_n(x,x)}.$$
Alternatively, $\KK_n(x,x)$ can be replaced by its asymptotic distribution to get
	$$\Omega = x + {(-s,s) \over n \psi(x)},$$
where $\dkf\mu = \psi(x) \dx$ is the equilibrium measure of $V$.  For $x$ inside the support of $\mu$, this statistic approaches the sine kernel distribution.   We demonstrate this in \figref{DegenerateGap1} for the degenerate potential, showing that the rate in which the statistics approach universality depends on the magnitude of the equilibrium measure.

\Figuretwo{ExpEM}{ExpEdge}
	The equilibrium measure for $V(x) = \E^x - x$  (left) and the scaled gap statistic  for $n = 10, 20, 40$ and 80 (right).  The dashed line is the Tracy--Widom distribution ($n = \infty$).

We now turn our attention to edge statistics.  In the generic position (i.e., when the equilibrium measure has precisely square root decay at its right endpoint), the gap probability for
	$$\Omega = \pr(b + {s \over c n^{2/3}},\infty)$$
tends to the Tracy--Widom distribution; here $c$ is a constant associated with the equilibrium measure, see \secref{TWScalingConstant} for its precise definition and the numerical method for its calculation.   In \figref{ExpEM}, we plot the computed equilibrium measure for $V(x) = \E^x - x$ (computed as described in \secref{EM}), and its scaled edge statistic for increasing values of $n$.  While the finite statistics are clearly converging to the Tracy--Widom distribution, the rate of convergence is much slower than the convergence of bulk statistics where the density of the equilibrium measure is large.

\Remark
	There are several methods for calculating universality laws --- i.e., $n = \infty$ statistics --- including using their \Painleve\ transcendent representations,  see \cite{BornemannRMTDistributions} for an overview.  An additional approach based on RH problems is to represent, say, 
	$$\partial_s \log  \det(I - {\cal S}|_{L^2(-s,s)})$$
 as a \RHp.  This can be solved numerically for multiple choices of $s$, and the results integrated numerically, see \cite{SOClaeysHigherOrderTW} for examples in the degenerate case.  This will be accurate in the tails, whereas the Fredholm determinant representation that we use only achieves absolute accuracy.  However, we are not aware of similar \RHp s for finite $n$.

\Section{RHP}  Orthogonal polynomials.

	We wish to calculate the monic polynomials $\pi_0(x),\pi_1(x),\dots$  with respect to the measure
		$$ \E^{-n V(x)} \dx$$
supported on the real line.
	Consider the following RH problem:

\Problem{OP} \cite{FokasOP}
The function
	$$Y(z) =  \sopmatrix{\pi_n(z)& \CC{[\pi_n \E^{-n V}](z)}  \cr
					\gamma_{n-1} \pi_{n-1}(z) & \gamma_{n-1} \CC[\pi_{n-1} \E^{-n V}](z)}$$
where
	$$\gamma_{n-1} ={ 2 \pi \I} \br[ \int_{-1}^1 \pi_{n-1}(x) w(x) \dx]^{-1}$$  solves the RH problem
	$$Y_+ = Y_- \sopmatrix{1 & \E^{-n V(x)}\cr & 1}\qqand Y \sim \sopmatrix{z^n \cr & z^{-n}}$$

To apply the numerical method described in \secref{nmRHP}, we must transform the RH problem for $Y$ into a suitable form for numerical solution.    To accomplish this, we will transform $Y$ by representing it explicitly in terms of  new functions which satisfy the following properties:
\beginorderedlist
	\newitem $Y  \mapsto T$ so that $T \sim I$ at infinity.
	\newitem $T \mapsto S$ so that the oscillatory jumps of $T$ become exponential decaying jumps of $S$.
	\newitem $S \mapsto \Phi$ so that the jumps of $\Phi$ are localized and scaled.
\endorderedlist

\Subsection Equilibrium measures. \label{sec:EM}

Our first task is to remove the growth in $Y$ at $\infty$.  To accomplish this, we must compute a so-called $g$-function  associated with the equilibrium measure of $V$:

\Definition{EM}
	The {\it equilibrium measure} $\mu$ is the minimizer of
	$$\iint\log{1 \over \abs{x - y}} \dkf\mu(x) \dkf\mu(y) + \int V(x) \dkf\mu(x).$$

  In this section, we assume that the equilibrium measure of $V$ is supported on a single interval $(a,b)$; a sufficient condition is that $V$ is convex \cite{DeiftOrthogonalPolynomials}.   (We remark that the below procedure was adapted to the multiple interval case in \cite{SOEquilibriumMeasure}, and adapting our numerical procedure for computing orthogonal polynomials, and thence invariant ensemble statistics, to such cases would be straightforward.)   
  
  With the correct choice of $(a,b)$, there exists $g$ analytic off $(a,b)$ satisfying 
	$$g_+(x) + g_-(x) = V(x) - \ell \qfor a \leq x \leq b \qqand g(z) \sim \log z.$$
The derivative $\phi = g'$ is analytic off $(a,b)$ and satisfies
	$$\phi_+(x) + \phi_-(x) = V'(x) \qfor a \leq x \leq b \qqand \phi(z) \sim {1 \over z}.$$
Given a candidate $(a,b)$, we can describe all $\phi$ satisfying this property
\Theorem{phi}\cite{SOEquilibriumMeasure}
	Denote the affine map from $(a,b)$ to $(-1,1)$ as
	$$M_{(a,b)}(z) = {2 z - a -b \over b - a}.$$		
Suppose we have
	$$V'(M_{(a,b)}^{-1}(x)) = \sum_{k=0}^\infty V_k T_k(x),$$
where $T_k$ is the $k$th order Chebyshev polynomial of the second kind.  If
	$$\phi_+(x) + \phi_-(x) = V'(x) \qfor x \in (a,b) \qand \phi(\infty) = 0,$$
then there exists a $\chi$ such that
	$$\phi(z) = \sum_{k=0}^\infty V_k J_+^{-1}(M_{(a,b)}(z))^k- {V_k \over 2}  M_{(a,b)}(z) {b -a \over 2 \sqrt{z - b} \sqrt{z - a}} + \chi {b -a \over 2 \sqrt{z - b} \sqrt{z - a}}$$
for the inverse Joukowski transform
	$$J_+^{-1}(z) = z - \sqrt{z-1} \sqrt{z+1}.$$

\Sketch
	This theorem follows from Plemelj's lemma and the fact that
	$$T_k(x) = {J_\downarrow^{-1}(x)^k + J_\downarrow^{-1}(x)^{-k} \over 2},$$
where
	$$J_\downarrow^{-1}(x) = x - \I \sqrt{1 - x} \sqrt{1 + x} = \lim_{\epsilon \downarrow 0} J_+^{-1}(x + \I \epsilon).$$

\mqed

To achieve the desired properties, we want $\phi$ to be bounded:
	$$V_0 = 0\qqand \chi = 0.$$
We also want $\phi(z) \sim {1\over z}$:
	$${b - a \over 8} V_1 = 1.$$
These two conditions give us a function
	$$F(a,b) = \Vectt[V_0,(b-a) V_1 - 8]$$
for which we want to find a root.  We can calculate $V_0$ and $V_1$ to high accuracy using the trapezium rule applied to
	$$\int_{-1}^1 {V'(M^{-1}(x)) T_k(x) \over \sqrt{1 - x^2}} \dx = -2 \int_{-\pi}^\pi V'(M^{-1}(\cos \theta)) \cos k \theta \dkf\theta.$$
This calculation is trivially differentiable with respect to $a$ and $b$, hence we can easily apply Newton iteration to find a root of $F$.    Convexity ensures that this root  is unique \cite{SOEquilibriumMeasure}.

Once $(a,b)$ are computed,  we calculate $\phi(z)$ by using the discrete cosine transform to calculate the Chebyshev coefficients of $V'$.  We then have the equilibrium measure
	$$\dkf \mu = {\I \over 2 \pi} \br[\phi^+(x) - \phi^-(x)] \dx = {\sqrt{1 - M_{(a,b)}(x)^2} \over 2 \pi}  \sum_{k=1}^\infty V_k U_{k - 1}(M_{(a,b)}(x))   \dx$$
where $U_k$ are the Chebyshev polynomials of the second kind.

To calculate $g$, we compute an indefinite integral of $\phi$ \cite{SOEquilibriumMeasure}:
	$$\eeq{
		g(z) = \int^z \phi(z) \dz = {b - a \over 4} \Biggl[V_1\pr({{J_+^{-1}(M_{(a,b)}(z))^2 \over 2}  - \log J_+^{-1}(M_{(a,b)}(z))}) \cr
		& \qqqquad\qqqquad\qqqquad + \sum_{k=2}^\infty V_k \pr({J_+^{-1}(M_{(a,b)}(z))^{k + 1} \over k + 1} - {J_+^{-1}(M_{(a,b)}(z))^{k - 1} \over k - 1}) \Biggr].
		}$$
This formula was derived by mapping $J_+^{-1}(M_{(a,b)}(z))$ back to the unit circle, where it became a trivially integrable Laurent series.  Note that $g$ has a branch cut along $(-\infty,a)$ on which it satisfies:
	$$g_+(x) - g_-(x) = 2 \pi \I.$$
Choosing (arbitrarily) $x \in (a,b)$, we calculate 
	$$\ell = V(x) - g_+(x) - g_-(x).$$

The numerically calculated $g$ consists of approximating $V_k$ using the discrete Cosine transform and truncating the sum.  Due to analyticity, the errors in these computed coefficients are negligible, and the approximation of $g$ is uniformly accurate in the complex plane.  Hereafter, we treat the numerical $g$ and the true $g$  as equal.  

\Subsection Scaling constant for edge statistics.  \label{sec:TWScalingConstant}

	Associated with the equilibrium measure are the {\it Mhaskar--Rakhmanov--Saff numbers}.  We re-express the constant as stated in \cite{DeiftEdgeUniversality} in terms of constants that we have already calculated: the support of the equilibrium measure and its Chebyshev coefficients.   The equilibrium measure for the scaled potential
	$V(M_{(a,b)}^{-1}(x))$
has support $(-1,1)$.  Its equilibrium measure is 
\meeq{
	{M_{(a,b)}^{-1}}'(x) \psi(M_{(a,b)}^{-1}(x)) \dx = {b - a \over 2} \psi(M_{(a,b)}^{-1}(x)) \dx =  (b - a){\sqrt{1 - x^2} \over 4 \pi}  \sum_{k=1}^\infty V_k U_{k - 1}(x)   \dx \ccr
 = {\sqrt{1 - x^2} \over 2 \pi} h(x) \dx,
}
as in \cite[(3.3)]{DeiftEdgeUniversality}.  We define the constant
	$$\alpha =\pr({h(1)^2 \over 2})^{1/3} = \half \br[(b - a) \sum_{k=1}^\infty k V_k]^{2/3}$$
as in \cite[(3.10)]{DeiftEdgeUniversality}.  
The scaling constant is thus
	$$c = { 2 \alpha  \over b - a} =  (b - a)^{-1/3} \br[\sum_{k=1}^\infty k V_k]^{2/3}.$$

\Subsection Lensing the \RHp.

We can now rewrite $Y$ to normalize the behaviour at infinity:
		$$Y(z) =  \sopmatrix{\E^{n \ell \over 2} & \cr & \E^{-{n \ell \over 2}} } T(z) \sopmatrix{\E^{- n g} & \cr & \E^{n g}} \sopmatrix{\E^{-{n \ell \over 2}} & \cr & \E^{n \ell \over 2} } ,$$
so that $T \sim I$ and has a branch cut along the unit interval, on which it satisfies
	$$\eeq{
	T_+ 
		=  T_- \sopmatrix{\E^{n (g_- - g_+)}  & \E^{n (g_+ + g_- + \ell  - V)}    \cr & \E^{n (g_+ - g_-)}} \\
		=T_- \socases{
		\sopmatrix{1 & \E^{n (g_+ + g_- + \ell  - V)}    \cr & 1}   & x < a \hbox{ or } x > b \cr
		\sopmatrix{\E^{n (g_- - g_+)}  & 1    \cr & \E^{n (g_+ - g_-)}} & a < x < b.
		}
		}$$

\Figurew{Lensing}{.9\hsize}
	The jumps of $T$.

	We appeal to properties of equilibrium measures (see \cite{SaffPotential}) to assert that 
	$$g_+(x) + g_-(x) + \ell -V < 0$$
for $x < a$ and $x > b$, thus those contributions of the jump matrix are isolated around $a$ and $b$.  On the other hand, $g_+ - g_-$ is imaginary between $a$ and $b$, hence $\E^{\pm n (g_+ - g_-)}$ becomes increasingly oscillatory on $(a,b)$.  We wish to deform the \RHp\ into the complex plane to convert oscillations into exponential decay.   To accomplish this, we introduce the lensing as in \figref{Lensing}, where we rewrite $T$ by
	$$T(z) = S(z) \socases{
		\sopmatrix{1 & \cr \E^{n (V - \ell - 2 g)} & 1} & z \in \Sigma_+\cr
		\sopmatrix{1 & \cr  \E^{n (V - \ell - 2 g)} & 1} & z \in \Sigma_- \cr
			I& \hbox{otherwise.} 
				}$$
By substituting
	$$g_+ = V - g_- - \ell,$$
we see that the oscillations have been removed completely from $\supp \mu$: 
	\meeq{
		S_+ = T_+ \sopmatrix{1 & \cr -\E^{n (V - \ell - 2 g_+)} & 1}   = T_+ \sopmatrix{1 & \cr -\E^{n (g_- -  g_+)} & 1}   \ccr
					 = 
		T_- \sopmatrix{ \E^{n (g_- - g_+)} & 1 \cr & \E^{n (g_+ - g_-)}}  \sopmatrix{1 & \cr -\E^{n (g_- -  g_+)} & 1}	\ccr
			 = T_- \sopmatrix{ & 1 \cr -1  &  \E^{n (g_+ - g_-)} }  = S_- \sopmatrix{1 & \cr -\E^{n (V - \ell - 2 g_-)} & 1} \sopmatrix{ & 1 \cr -1  &  \E^{n (V-\ell -2 g_-)} }  \ccr
			 	= S_- \sopmatrix{ & 1 \cr -1 &}.
			 }
However, we have introduced new jumps on $\Gamma_\uparrow$ and $\Gamma_\downarrow$, on which
	$$S_+ = T_+ = T_- = S_- \sopmatrix{1 & \cr \E^{n (V - \ell - 2 g)} & 1}.$$

\subsection{Removing the connecting jump}  \label{sec:Remove}

	We have successfully converted oscillations to exponential decay.  However, to maintain accuracy of the numerical algorithm for large $n$, we must isolate the jumps to neighbourhoods of the endpoints $a$ and $b$.   Thus we require  a function which satisfies the following RH problem:
	$$N_+(x) = N_-(x) \sopmatrix{ & 1 \cr -1 &}\qqfor a < x < b\qqand N(\infty) = I.$$
The solution is \cite{DeiftOrthogonalPolynomials}
	$$N(z)  = {1 \over 2 \nu(z) } \sopmatrix{1 &  \I \cr
	- \I & 1 } + {\nu(z) \over 2}  \sopmatrix{1 &  - \I \cr
	 \I & 1 } 
	\qfor \nu(z) = \pr({z - b \over z - a})^{1/4};$$
i.e., $\nu(z)$ is a solution to
	$$\nu_+(x) = \I \nu_-(x)\qfor a < x < b \qqand \nu(\infty) = 1.$$

An issue with using $N$ as a parametrix is that it introduces singularities at $a$ and $b$, hence we also  introduce local parametrices to avoid these singularities.  In the event that the equilibrium measure $\psi(x)$ has exactly {\it square root decay} at the edges, asymptotically accurate local parametrices are known.  However, if the equilibrium measure has higher order decay (\ala\ the higher-order Tracy--Widom distributions \cite{HigherOrderTracyWidom}), the asymptotically accurate local parametrices are only known in terms of a \RHp.  

For numerical purposes, however, we do not need the parametrix to be asymptotically accurate: we achieve asymptotic accuracy by scaling the contours.  Thus we introduce the {\it trivially constructed} local parametrices which satisfy the jumps of $S$ in neighbourhoods of $a$ and $b$:  
	$$P_a(z) = \pr(\socases{
	\sopmatrix{1 \cr 1 & 1} & {\pi \over 3} < \arg(z-a) < \pi \cr
	\sopmatrix{1& -1 \cr 1 & } & -\pi < \arg(z-a) <-{  \pi \over 3}  \cr
	\sopmatrix{& -1 \cr 1 & } & -{  \pi \over 3}< \arg(z-a) <0  \cr	
	I & \hbox{otherwise}
	})\sopmatrix{ \E^{n (V - \ell - 2 g)} \cr & \E^{-n (V - \ell - 2 g)}}$$
and
	$$P_b(z) = \pr(\socases{
	\sopmatrix{1 \cr -1 & 1} & {2 \pi \over 3} < \arg(z-b) < \pi \cr
	\sopmatrix{& -1 \cr 1 & 1} & -\pi < \arg(z-b) <-{ 2 \pi \over 3}  \cr
	\sopmatrix{1 & -1 \cr  & 1} & -{ 2 \pi \over 3} < \arg(z-b) < 0   \cr	
	I & \hbox{otherwise}
	})\sopmatrix{ \E^{n (V - \ell - 2 g)} \cr & \E^{-n (V - \ell - 2 g)}}.$$

\Figurew{FinalJumps}{0.9\hsize}
	The jumps of $\Phi$.

We can now write
	$$S(z) = \Phi(z) \socases{ N(z)  & \abs{z - a} > r \hbox{ and } \abs{z - b} > r \cr 
						P_b(z) & \abs{z - b} < r \cr
						P_a(z) & \abs{z - a} < r}$$
The final RH problem for $\Phi$ satisfies the jumps depicted in \figref{FinalJumps}.  

In practice, we do not use infinite contours.  We truncate contours when the jump matrix is, to machine precision, the identity matrix.  In all cases we consider here, after proper deformations the jump matrices are $C^\infty$ smooth  and are exponentially decaying to the identity matrix for large $z$.  The truncation of contours can be rigorously justified by solving a `nearby' RH problem with truncated contours.  For a full discussion of this see \secref{Uniform} and, in particular, Lemma~\ref{lemma:contourtruncation} below.  We can then deform the remaining contours to be line segments connecting their endpoints.   The resulting jump contour consists only of affine transformations of the unit interval.

\Section{nmRHP}  Numerical solution of Riemann--Hilbert problems. \label{NumericalRHP}

\def\cop<#1><#2>{{\cal C}\left[#1;#2\right]}
\def\capn<#1><#2>{{\cal C}_m\left[#1;#2\right]}
\def\sol{U}
\def\soln{U_m}
\def\ops<#1><#2>{{\cal L}\left(#1,#2\right)}
\def\op<#1>{{\cal L}\left(#1\right)}
\def\jump{G}
\def\cauchy<#1>{{\cal C}_{#1}}
\def\cauchyp<#1>{{\cal C}^+_{#1}}
\def\cauchym<#1>{{\cal C}^-_{#1}}
\def\cauchypm<#1>{{\cal C}^\pm_{#1}}

We have reduced the orthogonal polynomial RH problem to the form
\begin{align}\label{RHP}
\Phi^+(z) = \Phi^-(z) \jump(z), ~~ z \in \Gamma, ~~ \Phi(\infty) = I,
\end{align}
where $\Gamma = \Gamma^1 \cup \cdots \cup \Gamma^L$ is a union of contours that are affine transformations of the unit interval: i.e.,
	$M^{i}(\Gamma^i) = (-1,1)$ for an affine transformation $M^i$.  
We use the notation $[\jump;\Gamma]$ to refer to this RH problem and $\Phi = [\jump;\Gamma]$ when $\Phi$ is the unique solution.  Assume this solution has the form
\begin{align*}
\Phi(z) = I + \cauchy<\Gamma> \sol,
\end{align*}
for some smooth function $\sol$.  The RH problem \eqref{RHP} is converted into an equivalent singular integral equation (SIE) by substituting this assumed form into \eqref{RHP}:
\begin{align}\label{RHP-subs}
I + \cauchyp<\Gamma> \sol = (I + \cauchym<\Gamma> \sol)\jump.
\end{align}
We use the operator identity \cite{DeiftOrthogonalPolynomials}
\begin{align*}
\cauchyp<\Gamma> - \cauchym<\Gamma> = I,
\end{align*}
to rewrite \eqref{RHP-subs}
\begin{align}
\sol - \cauchym<\Gamma> \sol (\jump - I) = \jump - I.  \label{eq:SIE}
\end{align}
It is well-known that the operators $\cauchypm<\Gamma>$ are bounded from $L^2(\Gamma)$ to itself for every contour we consider here.  We use the notation
\begin{align*}
\cop<G><\Gamma> \sol = \sol - \cauchym<\Gamma> \sol (\jump -I),
\end{align*}
which is a well-defined bounded linear operator on $L^2(\Gamma)$ provided $\jump \in L^\infty(\Gamma)$.

	Our numerical scheme consists of approximating $U$ by a finite-dimensional sum of  {\it mapped Chebyshev polynomials}.  In other words, for $x \in \Gamma$ we approximate $U(x) \approx U_{\vc m}(x)$, where we define
	$$U_{\vc m}(x) = U_{\vc m}^i(x) = \sum_{k = 0}^{m^{i}- 1} U_k^{i} T_k(M^{i}(x))\qqfor x \in \Gamma^i,$$
for as-of-yet unknown coefficients ${U_k^{i}} \in \C^{2 \times 2}$.  If we are given the coefficients, we can evaluate $\cop<G><\Gamma> U_{\vc m}$  {\it pointwise} by using an exact expression for the Cauchy transform of our basis:

\Proposition{cauchymap}\cite{SOHilbertTransform}
	$${\cal C}_{\Gamma^i}[T_k \circ {M^i}](z) = {\cal C}_{(-1,1)}T_k({M^i}(z))$$	

\Sketch
	Follows from Plemelj's lemma: 
	$$ {\cal C}_{(-1,1)}T_k({M^i}(\infty)) = {\cal C}_{(-1,1)}T_k(\infty) = 0$$
and
	$$\br[{\cal C}_{(-1,1)}^+ - {\cal C}_{(-1,1)}^-]T_k({M^i}(x)) = T_k(M^i(x)).$$
\mqed

\Theorem{cauchyT} \cite{SOHilbertTransform,SOPainleveII}
 Define
	$$\eeq{
	\psi_k(z)  
		     ={2 \over \I \pi} \socases{
		     \arctanh z & \hbox{for $k = 0$} \\
		     {z^{1 + 2 \floor{-k/2} + k}  \over (1 - z^2)(1 + 2 \floor{-k/2})}\,\, {}_2F_1\fpr({1,\ 1\atop {3 \over 2} + \floor{-k/2}  }; {{z^2 \over z^2-1
   }}) & \hbox{for $k < 0$} \cr
   \noalign{\vskip .1in}\cr
 z^k  (\arctanh z - \arctanh {z^{-1}})   & \cr
 \quad + {z^{k-1 - 2 \floor{(k+1)/2}}  \over (1 - z^{-2})(1 + 2 \floor{(k+1)/2})} \,\,{}_2F_1\fpr({1,\ 1\atop {3 \over 2} + \floor{(k+1)/2}  }; {{z^{-2} \over z^{-2}-1
   }}) & \hbox{for $k > 0$,} \cr   
   }
		}$$
%
where ${}_2F_1$ is the hypergeometric function \cite{DLMF}.  Then
	\meeq{
	\CC T_k(z) 
	=  -{1 \over 2} \br[\psi_k(\Tin) +\psi_{-k}(\Tin) ],
	}
where 
	$$\Tin = z - \sqrt{ z - 1} \sqrt{z + 1}$$
is again an inverse of the Joukowsky transform $J(z) = \half (z + z^{-1})$.

\Sketch
	This also follows from Plemelj's lemma after mapping to the unit circle:
	 $$T_k(J(z)) = {z^k +z^{-k} \over 2},$$
 and relating the Taylor series of $\arctanh z$ to the hypergeometric function.  
	
\mqed

We now choose the coefficients $U_k^i$ by enforcing \eqref{eq:SIE} to hold pointwise at a collection of $N= |\vc m| = m^1 + \cdots + m^\ell$ points.  In other words, we choose points $\set{z_1^i,\ldots,z_{ m_i}^i}$ lying on each $\Gamma_i$, and solve the $4N \times 4N$ linear system
	\mEq{IdealizedLinSys}{\cop<G><\Gamma> U_{\vc m}(z_k^i) = \jump(z_k^i) - I.}

We choose {\it mapped Chebyshev points} for the points:
	$$\set{z_1^i, \dots, z^i_{ m_i}}= \set{{M^i}^{-1}(-1),{M^i}^{-1}\pr({\cos \pi \br[1 - {1 \over m_i - 1}]}),\dots,{M^i}^{-1}(1)}.$$
  This means every junction point of $\Gamma$ is included in the collocation system, with multiplicity the number of contours emanating from the junction point.  We denote these repeated points by
	$$\set{\xi + 0 \E^{\I \theta_1},\dots,\xi + 0 \E^{\I \theta_L}}$$
where $\theta_1,\ldots,\theta_L$ are the angles in which the components of $\Gamma$ that include $\xi$ as a junction point emanate from  $\xi$.  
  But, as seen in \thref{cauchyT}, the Cauchy transform for our basis blows up at such points!  To overcome this discrepancy, we assume that the solution satisfies the {\it zero sum condition}:

\Definition{ZeroSum}
	$U_{\vc m}$ satisfies the zero sum condition if, at every junction point $\zeta$, it satisfies
	$$\sum p_i U_{\vc m}^i(\zeta) = 0$$
where $p_i = -1$ if the left endpoint of  $\Gamma^i$ is  $\zeta$, $p_i = 1$ if the right endpoint of  $\Gamma^i$ is  $\zeta$ and $p_i = 0$ if $\zeta$ is not an endpoint of $\Gamma^i$.  

We can define an alternate expression for the Cauchy transform at the junction points:

\Definition{tildeC}
For $z$ not an endpoint of $\Gamma^i$, 
	$$\tilde {\CC}_{\Gamma^i}[T_k \circ M^i](z) = \CC_{\Gamma^i}[T_k \circ M^i](z).$$
Otherwise, for $z_L^i$ the left endpoint of $\Gamma^i$ and $z_R^i$ the right endpoint, if $\theta \neq \theta_i$ define
	\meeq{
		\tilde {\CC}_{\Gamma^i}[T_k \circ M^i](z_L^i + 0 \E^{\I \theta}) = a_k^{\rm L} + \I r_k^{\rm L} \arg(- \E^{\I (\theta - \theta^i)}) \ccr
				\tilde {\CC}_{\Gamma^i}[T_k \circ M^i](z_R^i + 0 \E^{\I \theta}) = a_k^{\rm R} + \I r_k^{\rm R} \arg( \E^{\I (\theta - \theta^i)})
		}
for
\meeq{
	 a_k^{\rm L} = (-1)^k {\log 2 \over 2 \pi \I} + {(-1)^k \over \I \pi}\br[\mu_{k-1}(-1) + \mu_k(-1)] + r_k^{\rm L} \log \abs{{M^i}'}, r_k^{\rm L} =   - {(-1)^k \over 2 \pi \I} , \ccr
 a_k^{\rm R} =  - {\log 2 \over 2 \pi \I} + {1 \over \I \pi}\br[\mu_{k-1}(1) + \mu_k(1)] + r_k^{\rm R} \log \abs{{M^i}'},  r_k^{\rm R} =  {1 \over 2 \pi \I },	 
	 }
where
$$\mu_k(z) =\sum_{j=1}^{\floor{k + 1 \over 2}} {z^{2 j - 1} \over 2 j- 1}.$$
When $\theta = \theta_i$ define
	$$\tilde {\CC}_{\Gamma^i}^\pm[T_k \circ M^i](z_L^i + 0 \E^{\I \theta_i})$$
by the appropriate limits.  

The usefulness of this alternative definition is that it is equivalent to the standard Cauchy transform for functions which satisfy the zero sum condition:
	
\Lemma{ZeroSumTildeC}\cite{SORHFramework}
	If $U_{\vc m}$ satisfies the zero sum condition, then
	$$\tilde {\CC} U_{\vc m}(z) = {\CC} U_{\vc m}(z).$$

\Sketch
	Let $\zeta$ be a junction point of $\Gamma$.  From the asymptotic behaviour of $\arctanh z$, we see near $\zeta$ that if $\zeta$ is an endpoint of $\Gamma^i$ ,
	$$\CC_{\Gamma^i}[T_k \circ M^i](z) \sim -{ p_i \over 2 \pi \I} \log{\abs{z - z_L^i}} + C_{\theta,k}^i$$
where $\theta = \arg(z -\zeta)$ and $p_i$ is defined as in \df{ZeroSum}.  If $\zeta$ is not a junction point of $\Gamma^i$, then
	$$\CC_{\Gamma^i}[T_k \circ M^i](z) \sim \CC_{\Gamma^i}[T_k \circ M^i](\zeta) =: C_{\theta,k}^i.$$
(where $C_{\theta,k}^i$ is $\theta$ independent).  
    Thus,
	\meeq{
		\CC U_{\bf n}(z) = \sum_i \sum_k U_k^i \CC_{\Gamma^i}[T_k \circ M^i](z) \sim - \sum_i p_i U^i(\zeta) {1 \over 2 \pi \I} \log{\abs{z - z_l^i}} + \sum_i\sum_k  C_{\theta,k}^i\ccr
			= \sum_i\sum_k  C_{\theta,k}^i;
		}
since the zero sum condition ensures that
	$$\sum_i p_i U^i(\zeta) = 0.$$
 The remaining constant $C_{\theta,k}^i$, which we refer to as the {\it finite part}, are precisely the constants we  defined above.  
\mqed

Thus, assuming the coefficients $U_k^i$ are in the space so that $U_{\vc m}$ satisfies the zero sum condition, we can replace \eq{IdealizedLinSys} by
	\mEq{TrueLinSys}{\tilde \CC[G,\Gamma] U_{\vc m}(z_k^i) = \jump(z_k^i) - I.}
This is justified by the following:
\Lemma{ZeroSumSol}\cite{SORHFramework}
	If the linear system \eq{TrueLinSys} is nonsingular, then the calculated
	$U_{\vc m}$ satisfy the zero sum condition.  

\Sketch
	Let $\zeta$ be a junction point, and assume for simplicity that $p_i = 1$ or 0 and $\Gamma_i$ are ordered by increasing arguments $\theta_i$.  Define
	$$\Phi_i^\pm = I + \tilde \CC^\pm U_{\vc m}(\zeta + 0 \E^{\I \theta_i}),$$
and define 
	$$G_i = G(\zeta + 0 \E^{\I \theta_i})$$
(i.e., the limit of the jump along $\Gamma^i$).  The collocation system imposes 
	$$\Phi_i^+ = \Phi_i^- G_i.$$
But the definition of $\tilde \CC$ imposes that
	\meeq{
		\Phi_{i+1}^- = \Phi_i^+ + (\theta_{i+1} - \theta_i) S\qqand
		\Phi_1^- = \Phi_L^+ + (\theta_1 + 2 \pi - \theta_L) S
		}
for 
	$$S = -\sum_i p_i {U_i(0) \over 2 \pi \I}.$$
These equations give
	\meeq{
		\Phi_L^+ = \Phi_L^+ G_1 \cdots G_L + S\br[(\theta_1 + 2\pi -\theta_L) G_1 \cdots G_L + \sum_{i = 2}^L (\theta_i - \theta_{i-1}) G_i \cdots G_L ]
		\ccr
		= \Phi_L^+ + S\br[(\theta_1 + 2\pi -\theta_L) I + \sum_{i = 2}^L (\theta_i - \theta_{i-1}) G_i \cdots G_L]
		}
where we use the well-posedness of the RH problem:
	$$G_1 \cdots G_L = I.$$
If 
$$(\theta_1 + 2\pi -\theta_L) I + \sum_{i = 2}^L (\theta_i - \theta_{i-1}) G_i \cdots G_L$$
is nonsingular (the {\it nonsingular junction condition}), then $S = 0$, implying the zero sum condition.  

Now suppose the nonsingular junction condition is not satisfied, and we replace the condition in the collocation system that
	$$\Phi_L^+ = \Phi_L^- G_L,$$
with 
	$S = 0$.
We then have
	$$\Phi_{i+1}^- = \Phi_i^+ \qand  \Phi_1^- = \Phi_L^+.$$
Thus
	\meeq{
		\Phi_L^- G_L = \Phi_{L-1}^+ G_{L-1} G_L = \cdots = \Phi_{L-1}^+ G_1 \cdots G_{L-1} G_L \ccr
	=\Phi_L^+,
	}
and the removed condition is still satisfied.  
In other words, the two linear systems are {\it equivalent}.

\mqed

In conclusion, the fact that the linear system is nonsingular implies that the numerically constructed
	$$\Phi_{\vc m}(z) = I + \CC U_{\vc m}(z)$$
is analytic off $\Gamma$ and satisfies the correct jumps at the collocation points.  We can thus recover approximations to orthogonal polynomials from $\Phi_{\vc m}$ by undoing the transformations $Y \mapsto T \mapsto S \mapsto \Phi$.

We have one last task: we need to scale the contours so that the numerical algorithm remains accurate for all choices of $n$.

\Subsection Spaces.

We follow \cite{TrogdonSONNSD} and  interpret the operator defined by applying $\tilde{\mathcal C}_\Gamma$ and sampling the resulting function  at $\{z_k^i\}$ as mapping of  piecewise polynomials to piecewise polynomials.  The sampled function at the values $\{z_k^i\}$ can be identified with its unique piecewise-polynomial interpolant and we use $\mathcal I_{\vc m}$ to denote this interpolation operator.  Define $L_{m_i}^2(\Gamma^i)$ to be the space of matrices with entries being $m_i$th order polynomials.  When $\Gamma = \Gamma^1 \cup \cdots \cup \Gamma^L$ has intersection points we define
\begin{align*}
L_{\vc m}^2(\Gamma) = \bigoplus_{i=1}^L L_{m_i}^2(\Gamma^i).
\end{align*}
We define $L_{\vc m,z}^2(\Gamma)$ to be the closed subspace of $L_{\vc m}^2(\Gamma)$ consisting of functions that satisfy the zero sum condition.  Thus $\mathcal I_{\vc m} {\tilde {\mathcal C}}_\Gamma$ is a well-defined linear operator from $L_{\vc m,z}^2(\Gamma)$ to $L_{\vc m}^2(\Gamma)$.  As is mentioned in \cite{TrogdonSONNSD}, $\mathcal I_{\vc m}  {\tilde {\mathcal C}}[G;\Gamma]$ maps to a proper subspace of $L_{\vc m}^2(\Gamma)$.

For each component contour $\Gamma^i$ and $k \in \mathbb N^+$ we define $H^k(\Gamma^i)$ and $W^{k,\infty}(\Gamma^i)$ in the usual way \cite{TrogdonSONNSD}.  For the contour $\Gamma$
\begin{align*}
H^k(\Gamma) = \bigoplus_{i=1}^L H^k(\Gamma_i), ~~ W^{k,\infty}(\Gamma) = \bigoplus_{i=1}^L W^{k,\infty}(\Gamma^i).
\end{align*}
Define $H_z^k(\Gamma)$ to be the close subspace $H^k(\Gamma)$ consisting of functions whose $(k-1)$th-order derivatives each satisfy the zero sum condition.  Finally, for a Banach spaces $X$ we use $\mathcal L(X)$ to denote the Banach space of operators on $X$ with the induced operator norm.

\Section{Uniform}  Uniform approximation. 

In this section we describe how the convergence of the  numerical approximation of RH problems can be made uniform in a parameter.  We also refer to this uniformity as \emph{asymptotic stability} of the numerical method.  In the case of orthogonal polynomials, the relevant parameter is $n$, the degree of the polynomial.   We refer to the results of \cite{TrogdonSONNSD}.  We assume that we have a sequence of RH problems $[\jump_n;\Gamma_n]$ depending on the parameter $n$.  The theory of \cite{TrogdonSONNSD} requires some assumptions on $\Gamma_n$.  Assume
        $$ \Gamma_n = \Omega^1_n \cup \cdots \cup \Omega_n^l, $$
where $\{\Omega_n^j\}_{j=1}^l$ are mutually disjoint and have the form
        $$ \Omega^j_n = \alpha_n^j \Omega^j + \beta_n^j. $$
We assume $\Gamma_n$ is the disjoint union of contours, each of which is an affine transformation of a fixed contour.  Once we have this separation of $\Gamma_n$ we attempt to solve the RH problem $[\jump_n;\Gamma_n]$ in an iterative way.  We define the restricted jumps $\jump_n^j = \jump_n|_{\Omega_n^j}$ and the jumps after variable change $H_n^j(k) = \jump_n^j(\alpha_n^j k + \beta_n^j)$, $k \in \Omega_j$.  The following result can be found in \cite{TrogdonSONNSD}.  For notational simplicity we suppress the dependence on $n$.  But it is important to note in the general case every function, but no domain, will depend on $n$.

\begin{lemma}[Scaled and shifted solver] \label{lemma:scaledsolver}

Assume $\tilde \Phi_1 = [H^1;\Omega_1]$ and define $\Phi_1 = \tilde \Phi_1\left(\frac{z-\beta^1}{\alpha^1}\right)$.  Furthermore, for each $j = 2,\ldots,l$ define $\Phi_{i,j}(k) = \Phi_i(\alpha^j k + \beta^j)$ and set
\begin{align*}
\tilde \Phi_j = \left[ \Phi_{j-1,j} \cdots \Phi_{1,j} H_j \Phi_{1,j}^{-1} \cdots \Phi_{j-1,j}^{-1}; \Omega_j\right], ~~ \Phi_j(z) = \tilde \Phi_j\left(\frac{z-\beta^j}{\alpha^j}\right).
\end{align*}
Then $\Phi = \Phi_1 \cdots \Phi_l$ solves $[\jump_n;\Gamma_n]$.

\end{lemma}

This lemma states that we can treat each disjoint contour separately.  We first solve a RH problem on one contour and modify the remaining jumps with the solution.  This process is repeated until all contours are taken account of.   

We use the following rule of thumb to determine the proper scalings $\alpha_n^j$:

\begin{assumption} \label{scalings} If the jump matrix $G$ has a factor
  $e^{n \theta}$ and $\beta^j$ corresponds to a $q$th order
  stationary point  (\emph{i.e.}, $\theta(z) \backsim
  C(z-\beta^j)^q$), then the scaling which achieves asymptotic
  stability is (a constant multiple of) $\alpha_n^j = n^{-1/q}$.
\end{assumption}

\Figure{DegenerateOmega}
	The pre-scaled $\Omega^0$ used for non-degenerate endpoints and the pre-scaled $\Omega^1$ used for first order degenerate endpoints.  
 
 In the case of a non-degenerate equilibrium measure, $g(z) \sim c_a (z-a)^{3/2}$ and $g(z) \sim c_b (z-b)^{3/2}$;  we thus scale like $n^{-2/3}$:
 	$$\Omega_n^1 = -n^{-2/3} \Omega^0 + a \qqand \Omega_n^2 = n^{-2/3} \Omega^0 + b,$$
where $\Omega^0$ is depicted in \figref{DegenerateOmega}, and the angle of the contours are chosen to match the direction of steepest descent.  
 In the first order degenerate case (eg., $V(x) = {x^2 \over 5} - {4 \over 15} x^3 + {x^4 \over 20} + {8 \over 5} x$), $g(z) \sim c_b (z-b)^{7/2}$ and so we scale like $n^{-7/2}$ at the degenerate endpoint:  
 	$$\Omega_n^1 = n^{-2/3} \Omega^0 + a \qqand \Omega_n^2 = n^{-7/2} \Omega^1 + b,$$
where $\Omega^1$ is depicted in \figref{DegenerateOmega} (the angle is sharper to attach to the new direction of steepest descent).  Higher order degenerate equilibrium measures will require higher order scalings, but this can be determined systematically by investigating the number of vanishing derivatives of the equilibrium measure.   

	This is the final form of the RH problem that we used in the numerical calculations of \secref{RMT}.  The remainder of the paper is concerned with proving that this scaled and shifted RH problem achieves {\it asymptotic accuracy}, i.e., the error does not grow as $n$ becomes large.  
 
 \Subsection Conditions for uniform approximation.

 A significant question is whether each of these smaller RH problems is solvable.  From a practical numerical standpoint this possible issue does not seem to affect the conditioning of the method.  From a theoretical standpoint this question is settled for large $n$ in \cite{TrogdonSONNSD} provided $\alpha^j_n \rightarrow 0$ for all $j$ as $n \rightarrow \infty$ with some mild restrictions on $\beta_n^j$.

\begin{assumption}
Assume that the jump matrix $G$ is $C^\infty$ when restricted to each component $\Gamma^i$ of $\Gamma$ and decays to the identity matrix faster than any polynomial at each isolated endpoint of $\Gamma$ and at $\infty$ if $\infty \in \Gamma$.
\end{assumption}
This is true in all cases we consider here.  The following lemma is proved in \cite{TrogdonSONNSD}.

\begin{lemma}[Contour truncation]\label{lemma:contourtruncation}
For every $\epsilon >0$ there exists an matrix-valued function $G_\epsilon$ and a bounded contour $\Gamma_\epsilon$ such that
\begin{itemize}
\item $G_\epsilon  = I$ on $\Gamma\setminus \Gamma_\epsilon$,
\item $\|G_\epsilon - G\|_{L^2(\Gamma) \cap L^\infty(\Gamma)} < \epsilon$, and
\item $\|\mathcal C[G;\Gamma] - \mathcal C[G_\epsilon;\Gamma_\epsilon]\|_{\mathcal L(L^2(\Gamma))} < \epsilon \|\mathcal C^-_\Gamma\|_{\mathcal L(L^2(\Gamma))}.$
\end{itemize}
\end{lemma}

Note that when the jump matrix $G$ is the identity matrix then the solution of the RH problem is analytic across the jump.  In practice we truncate infinite contours to finite contours when the jump matrix is within machine precision of the identity matrix.  The lemma justifies this process and we always assume $\Gamma$ is bounded.

The following theorem is the fundamental result of \cite{TrogdonSONNSD} and gives the required tools to address the accuracy of the Riemann--Hilbert numerical methods for orthogonal polynomials for arbitrarily large $n$:

\begin{theorem} \label{th:direct}  Assume
\begin{itemize}
\item $\mathcal C[H^j_n, \Omega^j]^{-1}$ exists and the norm $\|\mathcal C[H^j_n,\Omega^j]^{-1}\|_{\mathcal L(L^2(\Omega^j))} \leq C$ for all $j$ and $n$,
\item $\|H_n^j\|_{W^{k,\infty}(\Omega^j)} \leq C$ for all $j$ and $n$, and
\item $\alpha_n^j \goto 0$ as $n \goto \infty$.
\end{itemize}
Then for $n$ sufficiently large
\begin{itemize}
\item The algorithm of Lemma~\ref{lemma:scaledsolver} has solutions at each stage,
\item The approximation $U_{n,m_j}^j$ of $U_n^j$, the solution of the SIE at stage $j$ in the algorithm of Lemma~\ref{lemma:scaledsolver} converges in $L^2$ norm, uniformly in $n$ provided  $m_i \goto \infty$ for all $i\leq j$.
\end{itemize}
\end{theorem}

The theorem states that if the contours $\Gamma_n^j$ all have decaying measure then local boundedness properties on each of the contours can be made global for $n$ large.  As we will see below, bounding the $W^{k,\infty}$ norms of the matrices $H_n^j$ is often straightforward and the boundedness properties of the inverse operator follows from the asymptotic analysis of the RH problem.

\begin{remark}
If $\Gamma_n = \Omega_n^1$ consists of just one scaled contour then the restriction that $\alpha_n^1 \goto 0$ can be removed due to the fact that $z = \alpha_n^1 k + \beta_n^1$ is a conformal change of variables for the whole problem and this leaves the Cauchy integral operators invariant.
\end{remark}

\begin{remark}
Similar results hold when the bounds in \thref{direct} are known for a `nearby' RH problem.  In this case bounds on the nearby RH problem give slightly weaker convergence properties that can still be seen to be uniform in an appropriate sense \cite{TrogdonSONNSD}.
\end{remark}

\subsection{The classical Airy parametrix}\label{sec:AiryParametrix}


In this section we present the deformation and asymptotic solution of the RH problem that is performed in the asymptotic analysis of the RH problem for orthogonal polynomials.  The results from this section can be found in \cite{DeiftOrthogonalPolynomials}.  For brevity of presentation in this section we deal with potentials of the form $V(x) = x^{2m}$.  For the asymptotic analysis and deformations in the more case of $V(x)$  polynomial see \cite{DeiftWeights1,DeiftWeights2,DeiftWeights3,DeiftWeights4}.

A sectionally analytic, matrix-valued function $\hat \Phi$ is constructed explicitly out of the Airy function $\Ai(s)$ and its derivative such that $T \hat \Phi^{-1} \goto I$ as $n \goto \infty$ where $T$ is the solution of the original but deformed RH problem.   The RH problem for the error $E = T \hat \Phi^{-1}$ has smooth solutions and is a near identity RH problem in the sense that the associated singular integral operator is expressed in the form $I - K_n$ with $\|K_n\|_{L^2} \goto 0$ as $n \goto \infty$.   Thus $E$ can be computed via a Neumann series for sufficiently large $n$.

The deformation proceeds much in the same way as \secref{Remove}, except we replace $P_a$ and $P_b$ with new functions $\psi_a$ and $\psi_b$ that are constructed out of the Airy function.  We now construct these functions.  As an intermediate step, define
\begin{align*}
\Psi(s) &= \left\{ \begin{array}{lr}
\left(\begin{array}{cc}
\Ai(s) & \Ai(\omega^2 s) \\ \Ai'(s) & \omega^2\Ai(\omega^2s) \end{array} \right) e^{-i \frac{\pi}{6} \sigma_3} &  0 < \arg s < \frac{2\pi}{3}\\
\left( \begin{array}{cc}
\Ai(s) & \Ai(\omega^2s) \\ \Ai'(s)  & \omega^2 \Ai'(\omega^2s) \end{array}\right) e^{-i \frac{\pi}{6}} & \frac{2\pi}{3} < \arg s < \pi\\
\left(\begin{array}{cc}
\Ai(s) & \Ai(\omega^2 s) \\ \Ai'(s) & \omega^2 \Ai'(\omega^2s) \end{array}\right) e^{-i \frac{\pi}{6} \sigma_3} \left(\begin{array}{cc} 1 & \\-1 & 1 \end{array} \right) & \pi < \arg s < \frac{4\pi}{3}\\
\left(\begin{array}{cc}
\Ai(s) & -\omega^2\Ai(\omega s) \\ \Ai'(s) & -\Ai'(\omega s) \end{array}\right) e^{-i \frac{\pi}{6} \sigma_3} \left(\begin{array}{cc} 1 & \\1 & 1 \end{array} \right) & \frac{4 \pi}{3} < \arg s < 2 \pi\\
\end{array} \right.\\
\omega &= e^{\frac{2 \pi i}{3}}.
\end{align*}

\begin{figure}[ht]
\centering
\includegraphics[width=.6\linewidth]{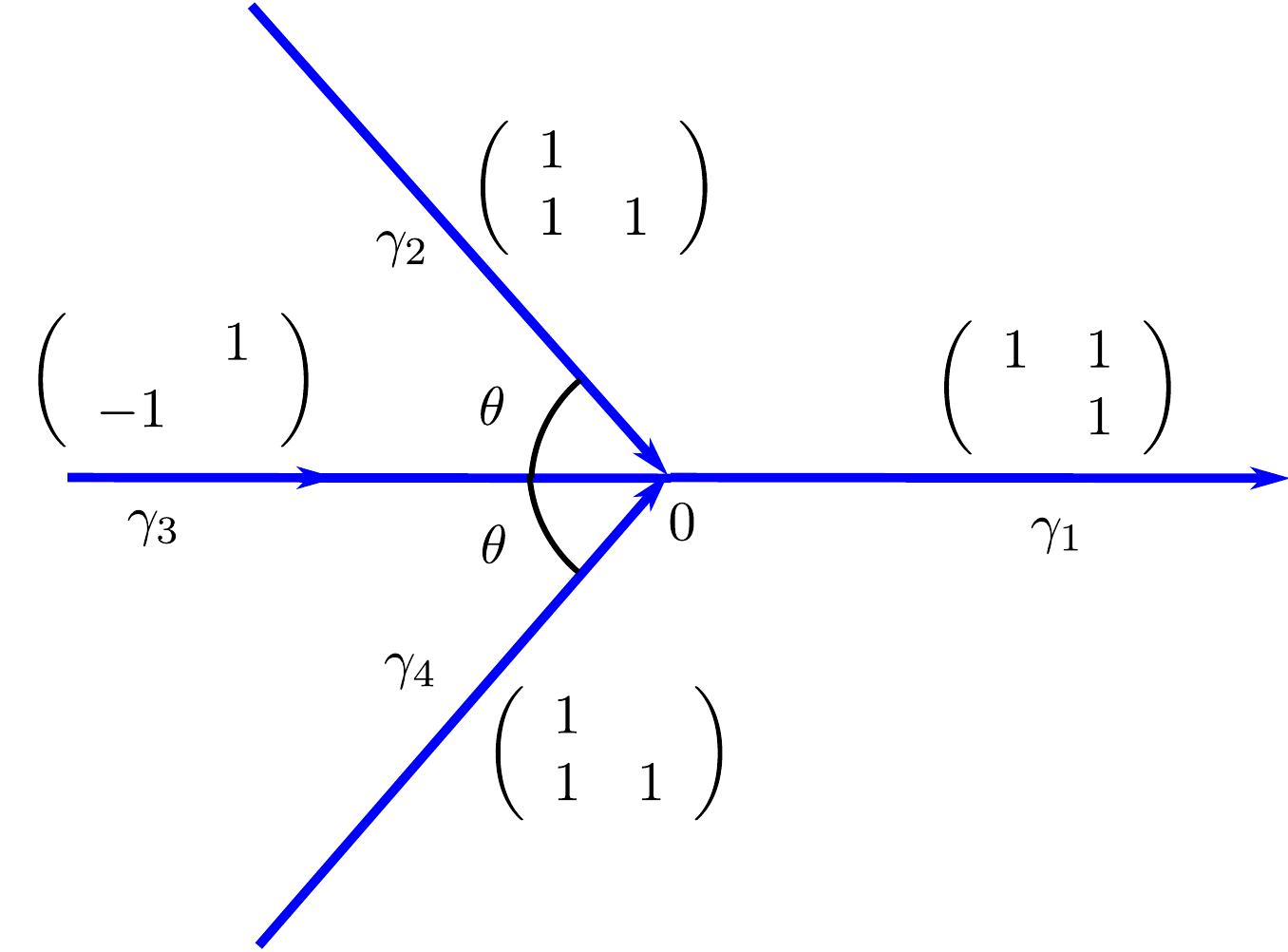}
\caption{The jump contours for $\Psi$ with jump matrices.  We include $\theta > 0$ in the figure for concreteness but its exact value is not needed below. \label{SimpleJumps}}
\end{figure}

The relations
\begin{align*}
\Ai(s) + \omega \Ai(\omega s) + \omega^2 \Ai(\omega^2 s) = 0,\\
\Ai'(s) + \omega^2 \Ai'(\omega s) + \omega \Ai'(\omega^2 s) = 0,
\end{align*}
can be used to show that $\Psi(s)$ satisfies the following jump conditions
\begin{align*}
\Psi^+(s) = \Psi^-(s) \left\{ \begin{array}{lc}
\left(\begin{array}{cc} 1 & 1 \\ & 1 \end{array} \right) & s \in \gamma_1\\
\left(\begin{array}{cc} 1 & \\ 1 & 1 \end{array} \right) & s \in \gamma_2\\
\left(\begin{array}{cc} & 1 \\ -1 &  \end{array} \right) & s \in \gamma_3\\
\left(\begin{array}{cc} 1 & \\ 1 & 1 \end{array} \right) & s \in \gamma_4
\end{array}\right. .
\end{align*}
See Figure~\ref{SimpleJumps} for $\gamma_i, ~ i =1,\ldots 4$.  

Since we only consider $V$ even in this section, the equilibrium measure is supported on a symmetric interval $[-a,a]$ for $a >0$.  Define
\begin{align*}
\Lambda (z) &= \frac{3}{2}\varphi(z) (z-a)^{-3/2}, ~~ \lambda(z) = (z-a)(\Lambda(z))^{2/3},\\
\varphi(z) &= \frac{1}{2}\left(V(z) - \ell\right) - g(z).
\end{align*}
It follows from the branching properties of $\varphi$ that $\Lambda$ and $\lambda$ are analytic in a neighbourhood of $a$.  Furthermore, since $\lambda(a) = 0$ and $\lambda'(a) = (\Lambda(a))^{2/3} \neq 0$ we use it as a conformal change of variables mapping a neighbourhood of $z = a$ into a neighbourhood of the origin. More precisely, fix an $\epsilon > 0$ and define $O_a = \lambda^{-1}(\{|z|< \epsilon\})$.

\begin{figure}[ht]
\centering
\subfigure[]{\includegraphics[width=.4\linewidth]{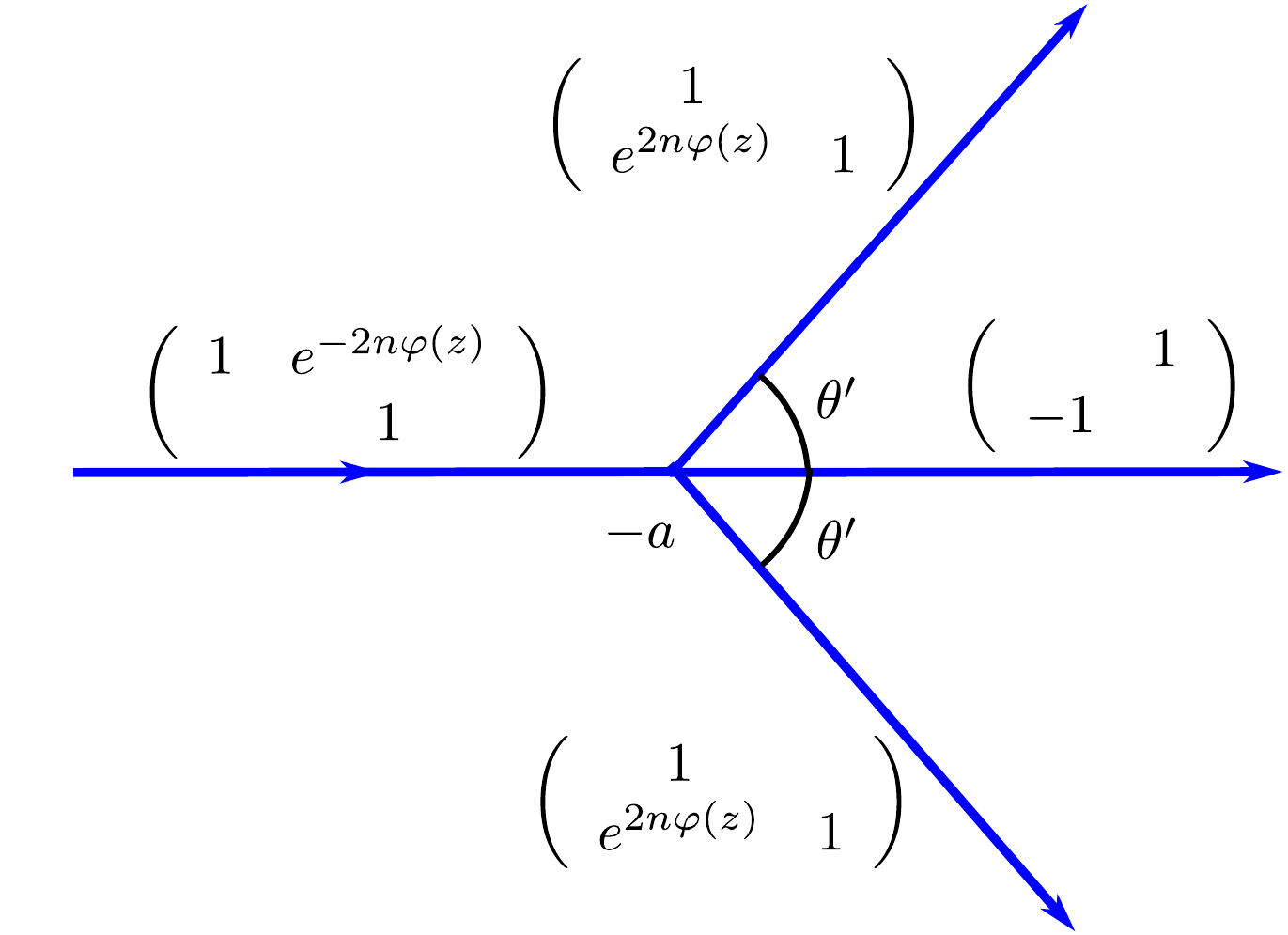}\label{PsinegaJumps}}
\subfigure[]{\includegraphics[width=.4\linewidth]{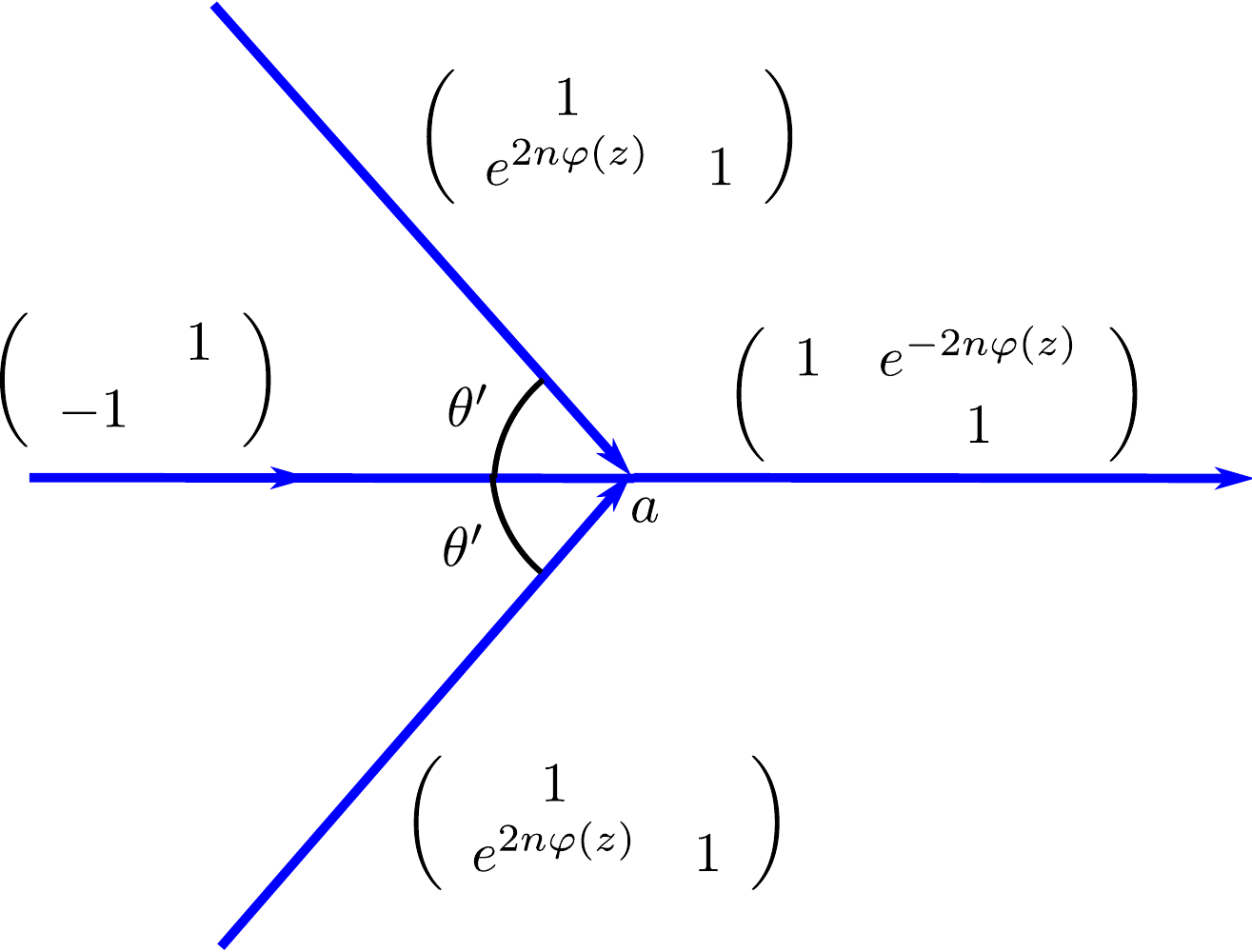}\label{PsiaJumps}}
\caption{The local parametrices near $z = a, -a$.  As above $\theta'>0$ is included for concreteness but its exact value is not needed. (a) The jump contours for $\psi_{-a}$ with jump matrices.  (b) The jump contours for $\psi_{a}$ with jump matrices. \label{PsiJumps}}
\end{figure}

Define
\begin{align*}
\psi_a(z) &= L(z)\Psi(n^{2/3} \lambda(z)) e^{n \varphi(z) \sigma_3},\\
L(z) &= \left(\begin{array}{cc} 1 & -1 \\ -i & -i \end{array} \right) \sqrt{\pi} e^{i\frac{\pi}{6}} n^{\sigma_3/6}((z+a)\Lambda^{2/3}(z))^{\sigma_3/4}.
\end{align*}
$\psi_a$ solves the local RH problem shown in Figure~\ref{PsiaJumps}.  The symmetry of $V(x)$ implies that $\psi_{-a}(z) = \sigma_3 \psi_{a}(-z) \sigma_3$ satisfies the jumps shown in Figure~\ref{PsinegaJumps}.  We are ready to define the full parametrix
\begin{align*}
\hat \Phi(z) = \left\{\begin{array}{lr}
\psi_{a}(z) & z \in O_a\\
\psi_{-a}(z) & z \in -O_a\\
N(z) & \text{otherwise}
\end{array} \right. .
\end{align*}

We will need a result concerning the asymptotics of the Airy function
\begin{align*}
\Ai(s) = \frac{1}{2 \sqrt{\pi}} s^{-1/4} e^{-\frac{2}{3} s^{3/2}} \left( 1 + \mathcal O\left(\frac{1}{s^{3/2}}\right)\right),\\
\Ai'(s) = -\frac{1}{2 \sqrt{\pi}} s^{3/4} e^{-\frac{2}{3} s^{3/2}} \left( 1 + \mathcal O\left(\frac{1}{s^{3/2}}\right)\right),
\end{align*}
as $s \goto \infty$ and $|\arg s| <\pi$.  These asymptotics, along with the definition of $\lambda(z)$, can be used to show
\begin{align}\label{bdry-converge}
\psi_a(z) N^{-1}(z) &= I + \mathcal O(n^{-1}), ~~ z \in \partial O_a,\\
\psi_{-a}(z) N^{-1}(z) &= I + \mathcal O(n^{-1}), ~~ z \in \partial O_{-a},
\end{align}
as $n \goto \infty$ uniformly in $z$ provided $O_a \cup O_{-a}$ is contained in a sufficiently narrow strip containing the real line.  See \cite{DeiftOrthogonalPolynomials} for the details.

We take the RH problem for $T$ in \figref{Lensing} and label $\partial O_a$ and $\partial O_{-a}$.  Note that without loss of generality we take $O_a$ and $O_{-a}$ to be open balls around $a$ and $-a$, respectively.  Analyticity allows us to deform any open, simply connected set containing $a$ or $-a$ to  a ball.


\begin{figure}[ht]
\centering
\includegraphics[width=.8\linewidth]{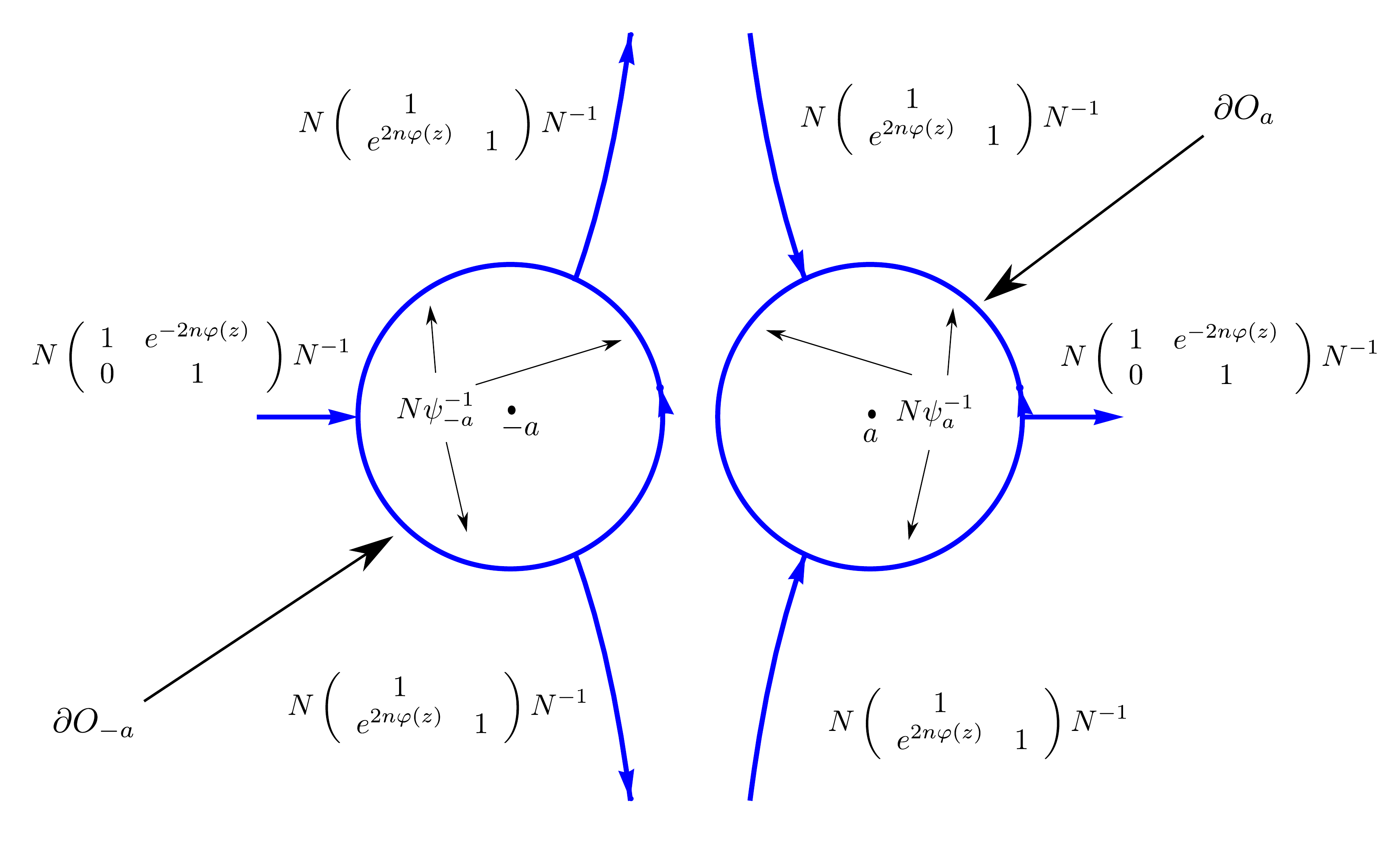}
\caption{The jump contours $\Omega$ for the error $E$.  The jump matrix $J$  for $E$ which is taken as the piecewise definition as shown. \label{FinalAiryJumps}}
\end{figure}

Since $\psi_a$ and $\psi_{-a}$ solve the RH problem locally in $O_a$ and $O_{-a}$, respectively, the function $E = T\hat \Psi^{-1}$ is analytic in  $O_a$ and $O_{-a}$.  See Figure~\ref{FinalAiryJumps} for the jump contour, $\Omega$, and jump matrix, $J$, for the RH problem for $E$. It is shown in \cite{DeiftOrthogonalPolynomials} using \eqref{bdry-converge} that the jump matrix for this RH problem tends uniformly to the identity matrix as $n \goto \infty$, again provided that all contours are in sufficiently small neighbourhood of the real line.  Thus
\begin{align*}
\|I - \cop<J><\Omega>\|_{\mathcal L(L^2(\Omega))} = \mathcal O(n^{-1}),
\end{align*}
and a Neumann series will produce the unique solution $u$ of $\cop<J><\Omega>u = J-I$.  $T$ is found via the expression
\begin{align*}
T(z) = (I + \mathcal C_{\Omega}u(z)) \hat \Psi(z).
\end{align*}

\Subsection Obtaining the bounds in \thref{direct}.

To apply \thref{direct} one has to first identify the correct scalings for the contours and second, establish bounds on the relevant operator norms and function derivatives.  

\subsubsection{The RH problem for $E$}\label{sec:Error}

In this case, we consider numerically solving the RH problem for $E$, rather than scaling and shifting the contours as we do in practice.  This simplifies the proof of uniform approximation considerably, at the expense of no longer allowing for degenerate potentials, and requiring significantly more knowledge in the construction of the RH problem. 

Take $\Gamma_n = \Omega$; that is, we do not scale the contour.  The near-identity nature of the RH problem allows us to avoid any scaling of the problem. Using the asymptotic expansions for the derivatives of Airy functions one can show that
\begin{align*}
\|J-I\|_{W^{k,\infty}(\Omega)\cap H^k(\Omega)} = \mathcal O(n^{-1}).
\end{align*}
Furthermore, the fact that $\|\cop<J><\Omega>^{-1}\|_{\mathcal L(L^2(\Omega))} < C$ follows easily from the Neumann series argument already given.  Thus we expect the numerical method to uniformly approximate solutions of this RH problem for small and arbitrarily large $n$.

\begin{figure}[ht]
\centering
\includegraphics[width=.5\linewidth]{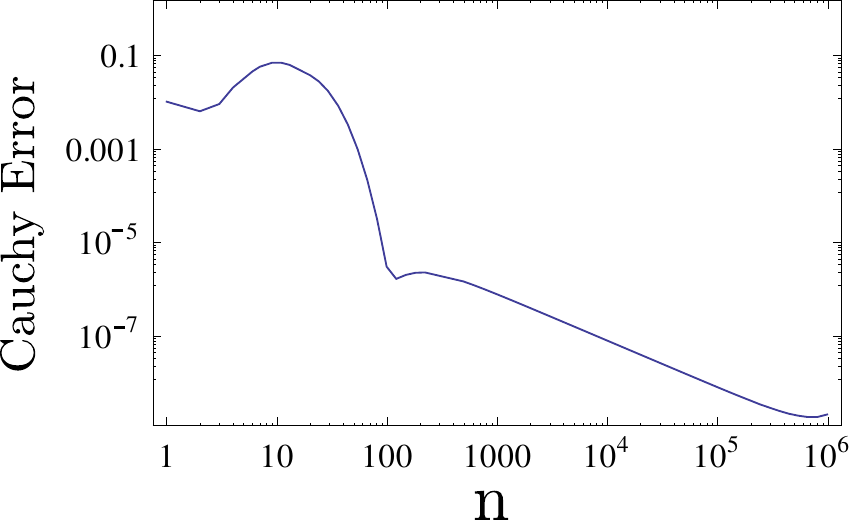}
\caption{\label{Airy-error}  The \emph{Cauchy error} between $U_{10}$ and $U_{20}$ as $n \goto \infty$.  This plot indicates that it takes fewer collocation points to approximate $E$ as $n$ increases.}
\end{figure}

To demonstrate the convergence properties of the solution for large $n$ we use the following procedure.  Let $U_{\vc m}$ denote the approximation of $u$ obtained using the numerical method for RH problems discussed above with $m =|\vc m|$ collocation points per contour.  When we break up $\Omega$ into both its non-self-intersecting components and components that can be represented by affine transformations of the unit interval we end up with 14 contours.  Thus, we use a total of $14 m$ collocation points.  We solve the RH problem with $m = 10$ and then again with $m = 20$.  We sample $U_{10}$ at each collocation point for $U_{20}$ and measure the maximum difference at these collocations points. We define this difference to be the \emph{Cauchy error}.  Figure~\ref{Airy-error} demonstrates that the error decreases as $n \goto \infty$.

\subsubsection{The RH problem for $\Phi$}\label{sec:unbounded}

The RH problem that we use in practice, $\Phi$, is of a fundamentally simpler form.  No additional special functions (e.g Airy functions) are needed and yet the contours are located away from the stationary points, $a$ and $b$ (we return here to allowing general potentials).  All deformations are performed by a reordering and analytic continuation of previously defined functions.

Assume we are in the non-degenerate case.  Represent 
\begin{align*}
\Gamma_n = \Omega_n^1 \cup \Omega_n^2,
\end{align*}
for
	$$\Omega_n^1 = n^{-2/3} \Omega^0 + a \qqand \Omega_n^2 = n^{-2/3} \Omega^0 + b.$$
%



%


Unfortunately, we have an issue with the jumps on these scaled contours: as $n \rightarrow \infty$, they  approach the unbounded singularities of $N(z)$, violating the conditions of \thref{direct}.  However, we can expand
	\meeq{
		N(a - z n^{-2/3}) = {n^{1/6} \over 2} \pr({b - a \over z})^{ {1 \over 4}} \sopmatrix{1 & -\I \cr  \I & 1}   + {n^{-1/6} \over 2 (b-a)} \pr({b - a \over z})^{ {3 \over 4}} \sopmatrix{1 & \I \cr  -\I & 1}  + \O(n^{-{1 \over 2}}), \ccr
		N(a - z n^{-2/3})^{-1} ={n^{1/6} \over 2} \pr({b - a \over z})^{ {1 \over 4}} \sopmatrix{1 & \I \cr  -\I & 1}   + {n^{-1/6} \over 2 (b-a)} \pr({b - a \over z})^{ {3 \over 4}} \sopmatrix{1 &- \I \cr  \I & 1} + \O(n^{-{1 \over 2}}).
		}
Letting 
	$$\bar N_{a,n} =n^{1/6} \sopmatrix{1 & -\I \cr  \I & 1}   + n^{-1/6} \sopmatrix{1 & \I \cr  -\I & 1}$$
we observe that 
	$$N(a + z n^{-2/3}) \bar N_{a,n}^{-1}\qqand \bar N_{a,n} N(a + z n^{-2/3})^{-1}$$
are uniformly bounded for $z$ restricted to an annulus around zero as $n \rightarrow \infty$. 

We thus remove the growth in the jumps by conjugating: let
	$$\Phi_1 = \bar N_{a,n} Q_1 \bar N_{a,n}^{-1}$$
outside $O_a$ (i.e., the simply connected region surrounding $a$) and 
	$$\Phi_1 = \bar N_{a,n} Q_1$$	
inside $O_a$.  The jumps  on $\partial O_a$ thus become: 
	$$Q_1^+ =  \bar N_{a,n}^{-1} \Phi_1^+ = \bar N_{a,n}^{-1} \Phi_1^- G = Q_1^- G \bar N_{a,n}$$
and, on the rest of $\Omega_n^2$,
	$$Q_1^+ =  \bar N_{a,n}^{-1} \Phi_1^+ \bar N_{a,n} = \bar N_{a,n}^{-1} \Phi_1^- G \bar N_{a,n}  = Q_1^- \bar N_{a,n}^{-1} G \bar N_{a,n},$$
so that $Q_1$ has bounded jumps.  

Once $Q_1$ and thence $\Phi_1$ are calculated, we need to bound the jump of $\Phi_2$, which is
	$$\Phi_1 G \Phi_1^{-1}.$$
Similar to before, we can find the two-term expansion of $N(b + z n^{-2/3})$ to find $\bar N_{b,n}$ to perform a second conjugation.  The asymptotic convergence of $\Phi_1$ to $I$ near $b$ ensures that $\bar N_{b,n}$ and $\Phi_1$ asymptotically commute.

	We can now appeal to \thref{direct} to show asymptotic stability.  The boundedness of the jumps can be verified directly.  To bound the inverse operators, we can use the boundedness of the parametrix of \secref{AiryParametrix} when the potential is non-degenerate.  
	In the degenerate case, we  we would need to use the analysis of the RH problem in \cite{HigherOrderTracyWidom}.  We omit the details here for brevity.



\Section{future}  Conclusion.

We presented a numerical method for computing statistics of unitary invariant ensembles, based on solving the associated Riemann--Hilbert problem numerically.  This required solving a nonlinear scalar Riemann--Hilbert problem to calculate the $g$ function associated with the equilibrium measure.  Scaling the contours appropriately resulted in a numerical method that remains accurate for large $n$, without knowledge of the local parametrices.  

	Our hope is that this framework will lead to a better understanding of the relationship between the potential $V$,  universality laws and finite $n$ statistics.



\appendix

\Section{PII} Computing a Hastings--McLeod Solution of the Painlev\'e II transcendent.  
\label{appendix:PII}

Here we focus on the (homogeneous) \PII\ ODE, it is as follows:
\begin{align}\label{PII}
u''(x) = x u(x) + 2 u^3(x).
\end{align}
(For brevity we refer to the homogeneous \PII\ simply as \Painleve\ II.) There
are many important applications of this equation:  the Tracy--Widom
distribution \cite{TracyWidom} from random matrix theory is written in
terms of the Hastings--McLeod solution \cite{HastingsMcLeod} and
asymptotic solutions to the Korteweg--de Vries and modified Korteweg--de
Vries equations can be written in terms of Ablowitz--Segur solutions
\cite{AblowitzSegurSolution}.  The aim of this section is to demonstrate
that the RH formulation can indeed be used effectively to compute
solutions to \PII, even in the asymptotic regime.

Solutions to differential equations such as \eqref{PII} are typically
defined by initial conditions: at a point $x$ we are given $u(x)$ and
$u'(x)$.  In the RH formulation, however, we do not specify initial
conditions.  Rather, the solution is specified by the \emph{Stokes'
  constants}; constants $s_1,s_2,s_3$ which satisfy the following
condition:

\begin{assumption}
\begin{align}\label{Stokes}
s_1 - s_2 + s_3 +s_1s_2s_3 = 0.
\end{align}
\end{assumption}

We will treat the Stokes' constants as given, as, in many applications they
arise naturally whilst initial conditions do not.  Given such
constants, we denote the associated solution to \eqref{PII}  by
\begin{align}
P_{\rm II}(s_1,s_2,s_3;z).
\end{align}
$P_{\rm II}$ and its derivative can be viewed as the special function which
map Stokes' constants to initial conditions.

\begin{figure}[ht]
\centering
\subfigure[]{\includegraphics[width=.4\linewidth]{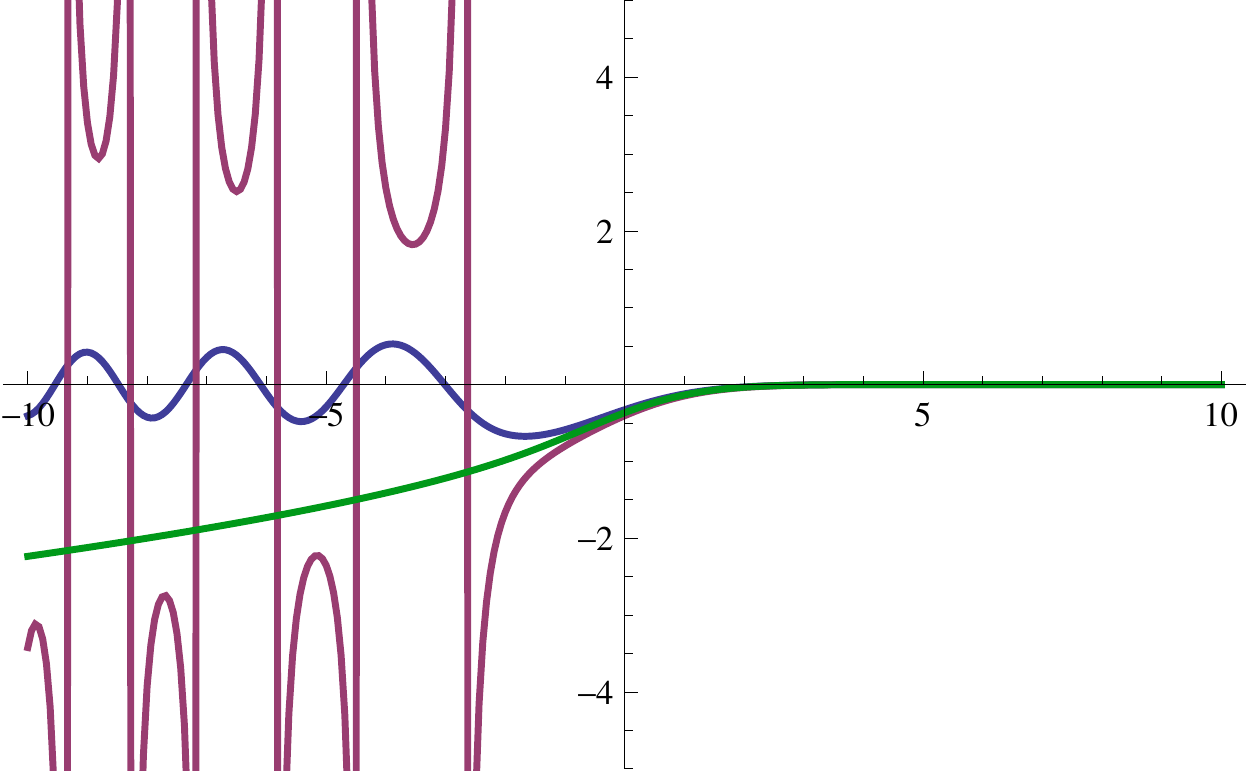}}
\hspace{.2in}
\subfigure[]{\includegraphics[width=.4\linewidth]{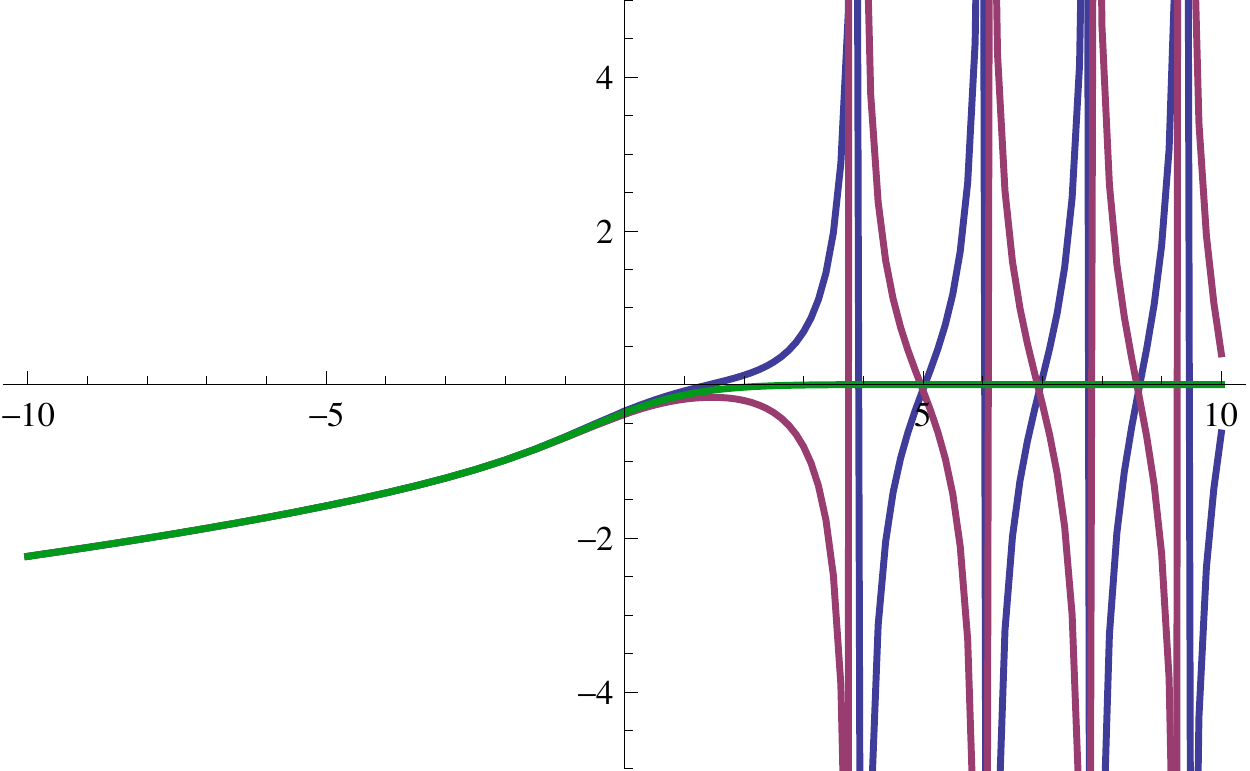}}
\caption{Solutions to \PII.  (a) Radically different solutions for $x
  <0$. (b) Radically different solutions for $x > 0$.}
\label{HastingsMcLeod}
\end{figure}

At first glance, computing solutions to \eqref{PII} appears trivial:
given initial conditions, simply use one's favorite time-stepping
algorithm, or better yet, input it into an ODE toolbox such as
\textsc{Matlab}'s \texttt{ode45} or  \textsc{Mathematica}'s
\texttt{NDSolve}.  Unfortunately, several difficulties immediately
become apparent.  In Figure \ref{HastingsMcLeod}, we plot several
solutions to \eqref{PII} (computed using the approach we are
advocating): the Hastings--McLeod solution and perturbations of the Hastings--McLeod solution.  Note that the solution is inherently unstable, and small perturbations cause oscillations --- which make
standard ODE solvers inefficient --- and poles --- which will
completely break  such ODE solvers (though this issue can be 
resolved using the methodology of \cite{FornbergPadePainleve}).



\begin{remark}
There are many other methods for computing the Tracy--Widom distribution
itself as well as the Hastings--McLeod solution
\cite{BornemannFredholm,BornemannRMTDistributions}, based on the
Fredholm determinant formulation or solving a boundary value
problem.  Moreover, accurate data values have been tabulated using
high precision arithmetic with a Taylor series method
\cite{PrahoferSpohnDataValues,PrahoferSpohnKPZ}.  However, we will see
that there is a whole family of solutions to \PII\ which exhibit
similar sensitivity to initial conditions, and thus a reliable,
general numerical method is needed even for this case.
\end{remark} 

\begin{figure}[ht]
\centering
\includegraphics[width=.5\linewidth]{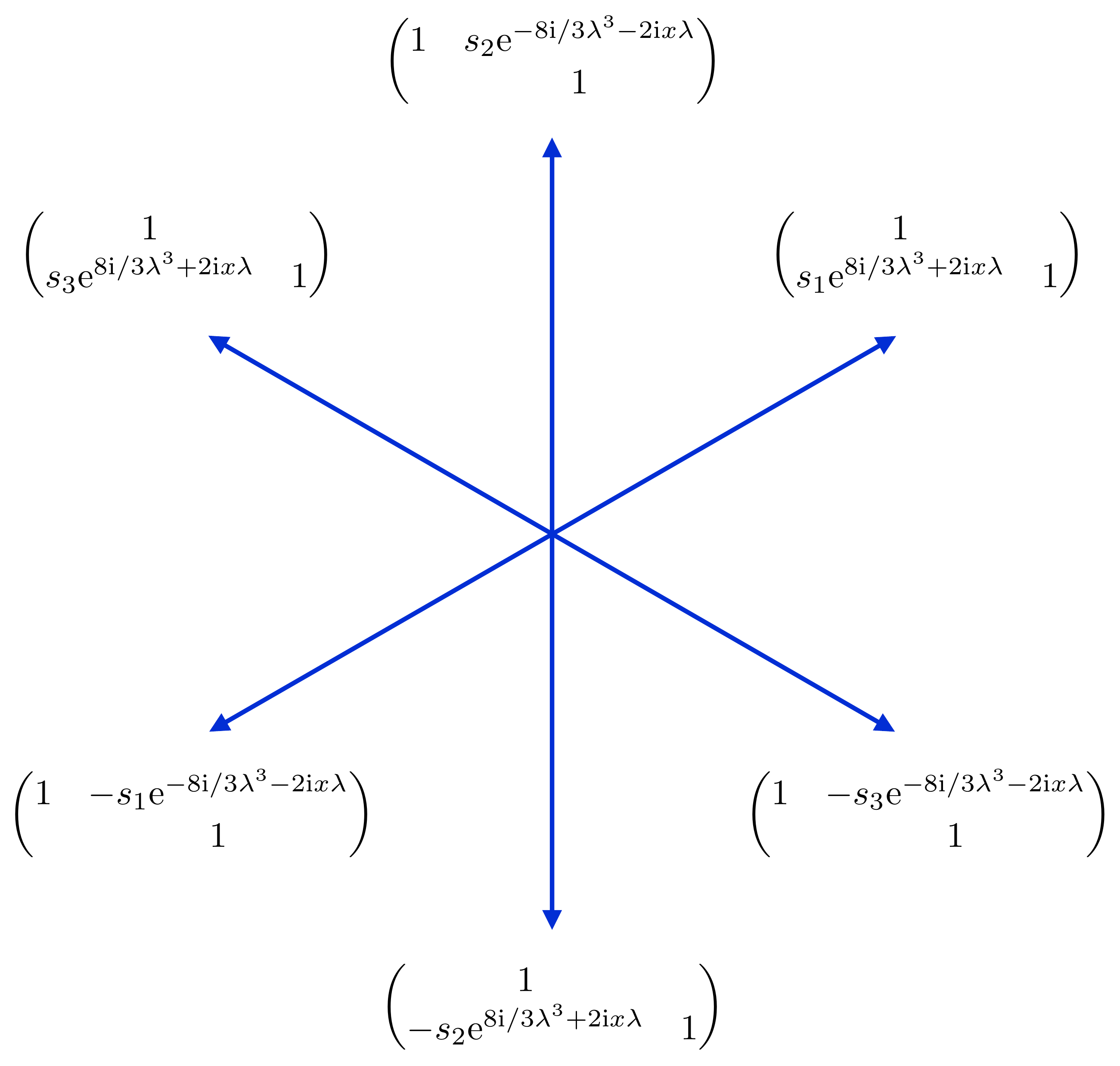}
\caption{\label{fig:UndeformedContour}The contour and jump matrix for the \PII\ RH problem.}
\end{figure}

Let $\Phi(x ; \lambda)$ solve the RH problem depicted in \figref{UndeformedContour}: let $\Gamma = \Gamma_1 \cup \cdots \cup \Gamma_6$ for $\Gamma_\kappa \{ s
e^{i\pi (\kappa/3-1/6)}: s \in \mathbb R^+\}$, \emph{i.e.}, $\Gamma$
consists of six rays emanating from the origin, as see in \figref{UndeformedContour}.  Then the jump matrix is defined by
$G(x;\lambda) = G_\kappa(x; \lambda)$ for $z \in \Gamma_\kappa$, where 
\begin{align*}
G_\kappa(x;\lambda) = G_\kappa(\lambda) = \begin{choices} \begin{mat} 1
    & s_\kappa e^{-i8/3 \lambda^3 -2i x \lambda} \\ & 1 \end{mat} \when
  \kappa \mbox{ even},\\
\begin{mat} 1 & \\  s_\kappa e^{i8/3 \lambda^3 +
    2ix\lambda} & 1 \end{mat} \when \kappa \mbox{ odd}. \end{choices}
\end{align*}
This is the RH problem which was solved numerically in
\cite{SOPainleveII}.
  We can recover the corresponding solution to \PII\ from $\Phi$ by \cite{FokasPainleve}
\begin{align*}
P_{\rm II}(s_1,s_2,s_3;x) = 2 \lim_{\lambda \goto \infty} \lambda \Phi(x;\lambda)_{12}.
\end{align*}




As $|x|$ becomes large, the jump matrices $G$ are increasingly
oscillatory.  We will combat this issue by deforming the contour so
that these oscillations be exponential decay.  To simplify this
procedure, we first rescale the RH problem.  Note that, if we let
$z = \sqrt{|x|} \lambda$, then the jump contour $\Gamma$ remains
unchanged, and
\begin{align*}
\Phi^+(z) = \Phi^+(\sqrt{|x|} \lambda ) = \Phi^-(\sqrt{|x|}
\lambda)G(\sqrt{|x|} \lambda) = \Phi^-(z) G(z),
\end{align*}
where $G(z) = G_\kappa(z)$ on $\Gamma_\kappa$ for
\begin{align*}
G_\kappa(z) = \begin{choices} \begin{mat} 1 & s_\kappa e^{-i|x|^{3/2}
      \theta(z)} \\ & 1 \end{mat} \when \kappa \mbox{ even},\\
\begin{mat} 1 & \\ s_\kappa e^{i|x|^{3/2}\theta(z)} & 1 \end{mat} \when
\kappa \mbox{ odd}, \end{choices}
\end{align*}
and
\begin{align*}
\theta(z) = \frac{2}{3} \left( 4z^3 + 2 e^{i \arg x} z\right).
\end{align*}
Then
\begin{align*}
P_{\rm II}(s_1,s_2,s_3;x) = 2i \lim_{\lambda \goto \infty} \lambda \Phi(x;\lambda)_{12} =
2i \sqrt{x} \lim_{\lambda \goto \infty} z \Phi(x;z)_{12}.
\end{align*}

\Subsection Positive $x$ with $s_1 = 0$.

We will now deform the RH problem for \PII\ so that numerics are
asymptotically stable for positive $x$.  We will see that the
deformation is extremely simple under the following assumption:

\begin{assumption} \label{s20} $s_2 = 0$ \end{assumption}

We remark that, unlike other deformations, the following deformation
can be easily extended to achieve asymptotic stability for $x$ in the
complex plane such that $-\frac{\pi}{3} < \arg x <
\frac{\pi}{6}$.

On the undeformed contour, the terms $e^{\pm i |x|^{3/2} \theta(z)}$ become
oscillatory as $|x|$ becomes large.  However, with the right choice of
curve $h(t)$, $e^{\pm i\theta(h(t))}$ has no oscillations; instead, it
decays exponentially fast as $t \goto \infty$.  But $h$ is precisely
the path of steepest descent, which passes through the stationary
points of $\theta$, \emph{i.e.}, the points where the derivative of
$\theta$ vanishes.  We readily find that
\begin{align*}
\theta'(z) = 2(4z^2+1),
\end{align*}
and the stationary points are  $z = \pm i/2$.

\begin{figure}[ht]
\centering
\includegraphics[width=.3\linewidth]{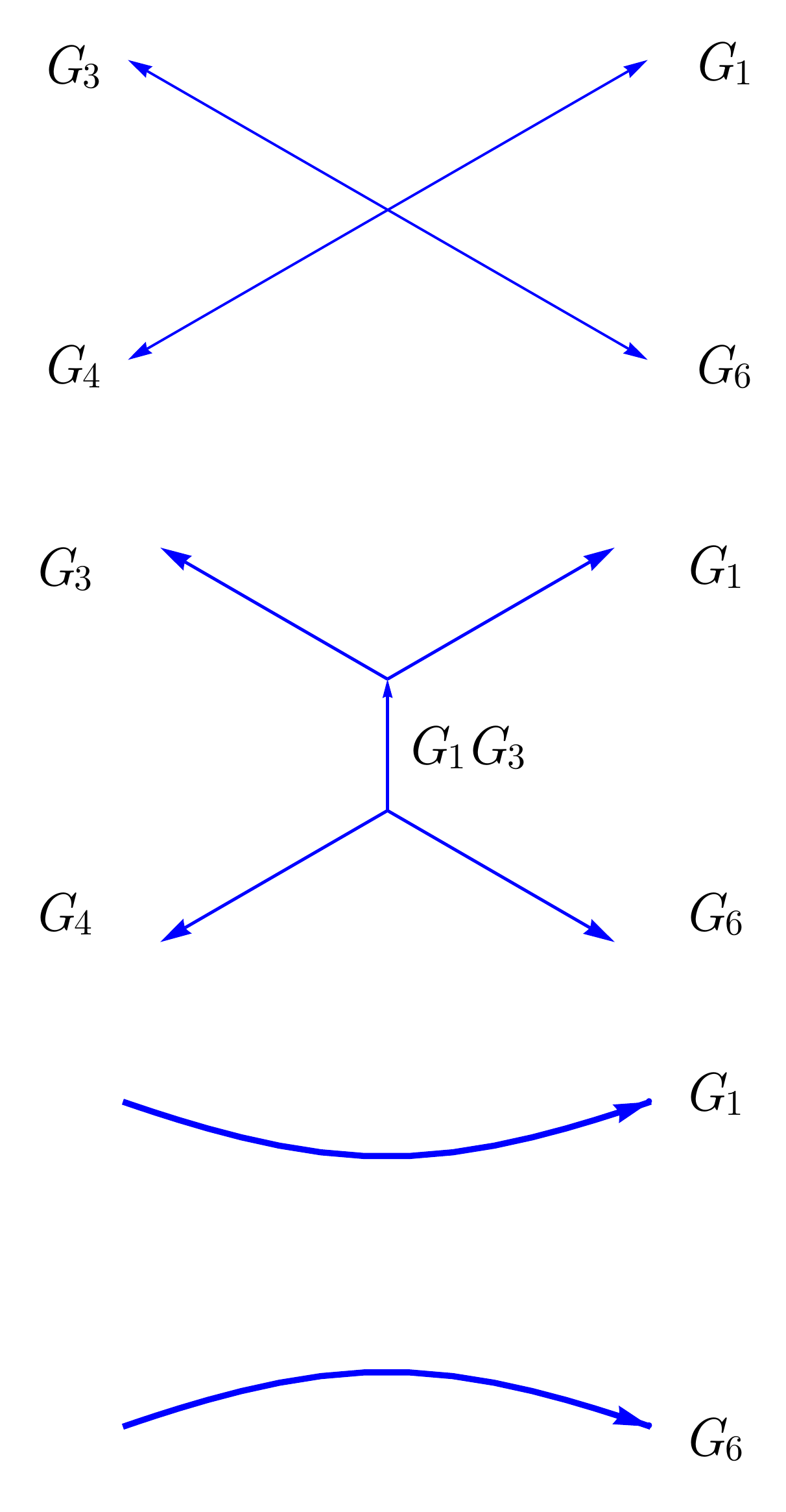}
\caption{\label{DeformedPositiveAS}Deforming the RH problem for positive $x$, with Assumption \ref{s20}.}
\end{figure}

We note that, since $G_2 = I$, when we deform $\Gamma_1$ and
$\Gamma_3$ through $i/2$ they become completely disjoint from
$\Gamma_4$ and $\Gamma_6$, which we then deform through $-i/2$.  We
also point out that $G_3^{-1} = G_1$ and $G_6^{-1} = G_4$; thus we can
reverse the orientation of $\Gamma_3$ and $\Gamma_4$, resulting in the
jump $G_1$ on the curve $\Gamma_{\uparrow}$ and $G_4$ on
$\Gamma_\downarrow$, as seen in Figure \ref{DeformedPositiveAS}.

Now recall that
\begin{align*}
\theta\left(\pm \frac{i}{2} \right) = \pm \frac{i}{3}.
\end{align*}
However, we  only have $\Gamma_\uparrow$ emanating from $i/2$, with
jump matrix
\begin{align*}
G_1 = \begin{mat} 1 & \\ s_1 e^{i|x|^{3/2} \theta(z)} & 1 \end{mat}.
\end{align*}
This is exponentially decaying to the identity along
$\Gamma_\uparrow$; as is $G_4$ along $\Gamma_\downarrow$.  We will
employ the approach of \secref{Uniform}. We first use Lemma~\ref{lemma:contourtruncation} to truncate the contours near the stationary point.  What remains is to determine what near means. Because $\theta$ behaves like
$\bigo\left(z \pm i/2\right)^2$ near the stationary points, Assumption \ref{scalings} implies that we should choose the shifting of $\beta_1 = i/2$ and $\beta_2 = -i/2$, the scalings $\alpha_1 = \alpha_2 = r |x|^{-3/4}$ and the canonical domains $\Omega_1 = \Omega_2 = [-1,1]$.  Here $r$ is chosen so that what is truncated is negligible in the sense of Lemma~\ref{lemma:contourtruncation}.  $G_6$ is similar.  The complete proof of asymptotic stability of the numerical method proceeds in a similar way as in Section~\ref{sec:Error}.

\Subsection Negative $x$ with $s_1 = - s_3 = \pm i$ and $s_2 = 0$.

We now develop deformations for the Hastings--McLeod solution for negative $x$, which corresponds to $s_1 = \pm \I$, $s_2 = 0$ and $s_3 = \mp \I$ \cite{FokasPainleve}.  We will realize numerical asymptotic stability in the aforementioned sense.

\begin{assumption} \label{s1s3} $s_1 = - s_3 = \pm i$ and $s_2 = 0$ \end{assumption}

\begin{figure}[ht]
\centering
\subfigure[]{\includegraphics[width=.45\linewidth]{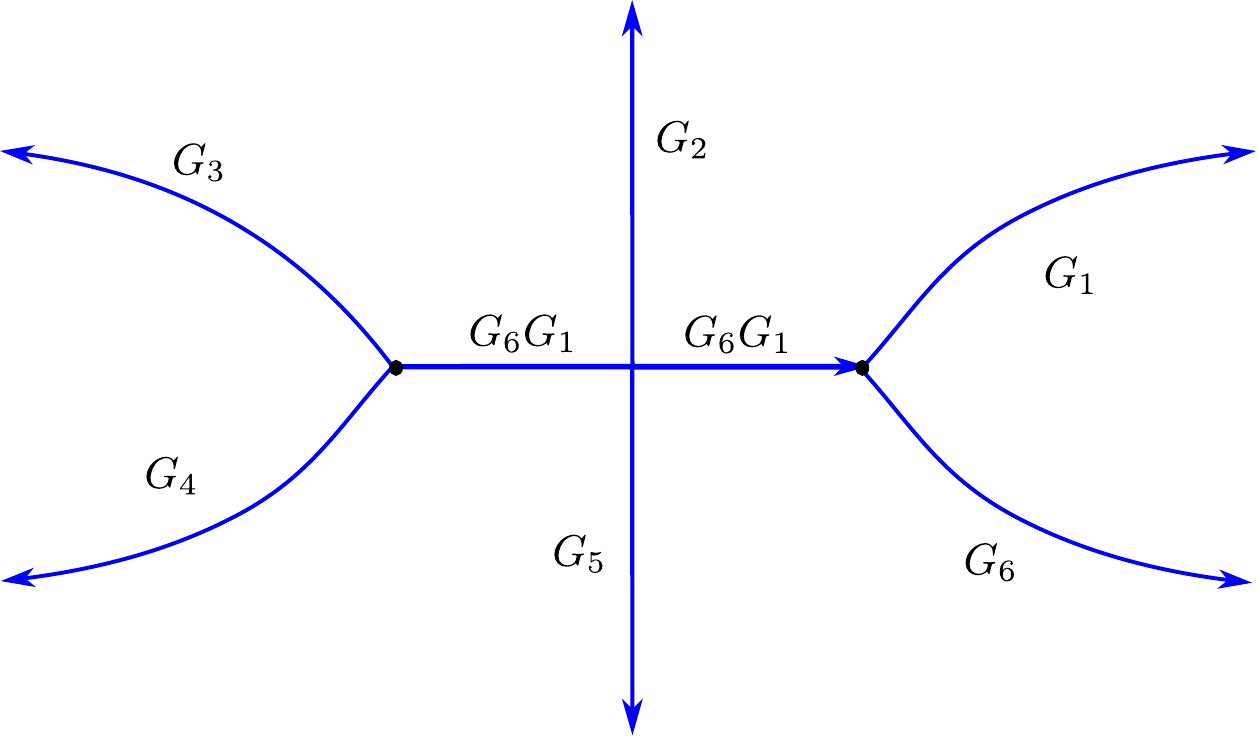}\label{HM1}}
\subfigure[]{\includegraphics[width=.45\linewidth]{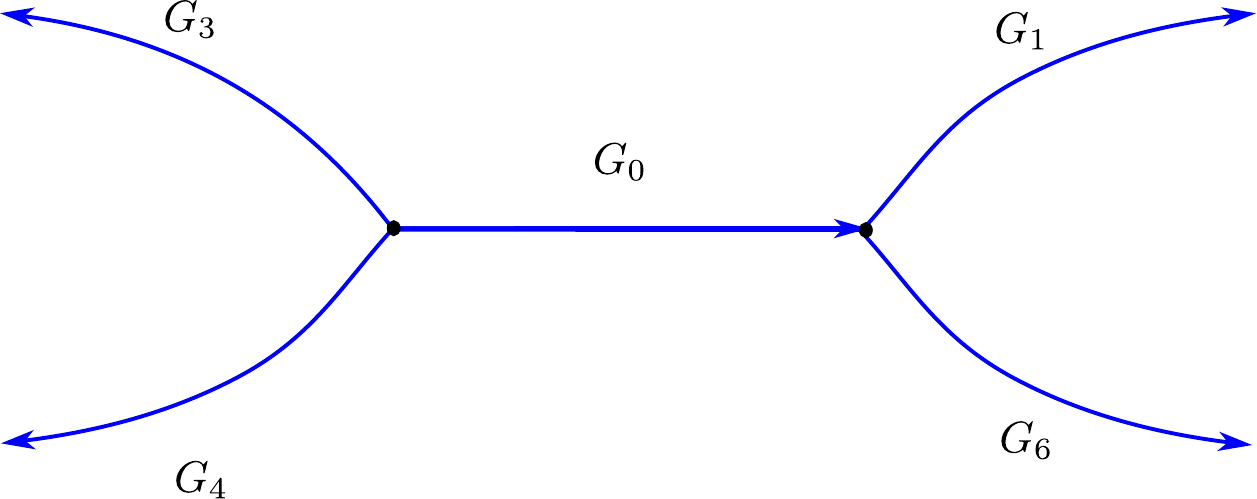}\label{HM2}}
\caption{\label{HM12}Deforming the RH problem for negative $x$, with Assumption \ref{s1s3}.  The black dots represent $\pm \alpha$.  (a) Initial deformation. (b) Simplification stemming from Assumption~\ref{s1s3}.}
\end{figure}

We begin by deforming the RH problem (\figref{UndeformedContour}) to the one shown in Figure~\ref{HM12}.  The horizontal contour extends from $-\alpha$ to $\alpha$ for $\alpha > 0$.  We will determine $\alpha$ below. Define
\begin{align*}
G_0 = G_6G_1 = \begin{mat} & s_1e^{-i |x|^{3/2} \theta(z)}  \\ s_1e^{i |x|^{3/2} \theta(z)} & 1 \end{mat}.
\end{align*}
Note that the assumption $s_2 = 0$ simplifies the form of the RH problem substantially, see Figure~\ref{HM2}.
We use an approach similar to that of the equilibrium measure to replace $\theta$ with a function possessing more desirable properties.  Define
\begin{align*}
\Theta(z) = e^{i |x|^{3/2} \frac{g(z)-\theta(z)}{2} \sigma_3}, ~~ \sigma_3 = \begin{mat} 1 &\\ & -1 \end{mat}, ~~ g(z) = (z^2-\alpha^2)^{3/2}.
\end{align*}
The branch cut for $g(z)$ is chosen along $[-\alpha,\alpha]$.  If we set $ \alpha = 1/\sqrt{2}$ the branch of $g$ can be chosen so that $g(z) -\theta(z) \sim \bigo(z^{-1})$.  Furthermore, $g_+(z) + g_-(z) = 0$ and $\text{Im} (g_-(z) - g_+(z)) > 0$ on $(-\alpha,\alpha)$.   Define $\hat G_i = \Theta^{-1}_-G_i \Theta_+$ and note that
\begin{align*}
\hat G_0(z) = \begin{mat}  & s_1 e^{-i|x|^{3/2} \frac{g_+(z) + g_-(z)}{2}} \\ 
s_1 e^{i|x|^{3/2} \frac{g_+(z) + g_-(z)}{2}} &  e^{i|x|^{3/2} \frac{g_-(z) - g_-(z)}{2}} \end{mat} =  \begin{mat}  & s_1  \\ 
s_1  &  e^{i|x|^{3/2} \frac{g_-(z) - g_-(z)}{2}} \end{mat}.
\end{align*}
As $x \goto -\infty$, $G_0$ tends to the matrix
\begin{align*}
J = \begin{mat} & s_1 \\ s_1 & \end{mat}.
\end{align*}

\newcommand{\hmout}{\Psi_{\text{HM}}^{\text{out}}}
\newcommand{\hma}{\Psi_{\text{HM}}^{\alpha}}
\newcommand{\hmna}{\Psi_{\text{HM}}^{-\alpha}}
\newcommand{\hm}{\Psi_{\text{HM}}}

The solution of the RH problem
\begin{align*}
\Psi^+(z) = \Psi^-(z)J, ~~ z \in [-\alpha,\alpha], ~~ \Psi(\infty) = I,
\end{align*}
is given by
\begin{align*}
\hmout(z) = \frac{1}{2} \begin{mat} \beta(z) + \beta(z)^{-1} & -i s_1(\beta(z) - \beta(z)^{-1})\\
-i s_1(\beta(z) - \beta(z)^{-1}) & \beta(z) + \beta(z)^{-1} \end{mat}, ~~\beta(z) = \left(\frac{z-\alpha}{z+\alpha}\right)^{1/4}.
\end{align*}
Here $\beta$ has a branch cut on $[-\alpha,\alpha]$ and satisfies $\beta(z) \goto 1$ as $z \goto \infty$.  It is clear that $(\hmout)_+ \hat G_0 (\hmout)^{-1}_- \goto I$ uniformly on every closed subinterval of $(-\alpha,\alpha)$.

\newcommand{\piot}{\frac{\pi}{3}}
\newcommand{\tpiot}{\frac{2\pi}{3}}

\begin{figure}[ht]\label{HM4}
\centering
\includegraphics[width=.6\linewidth]{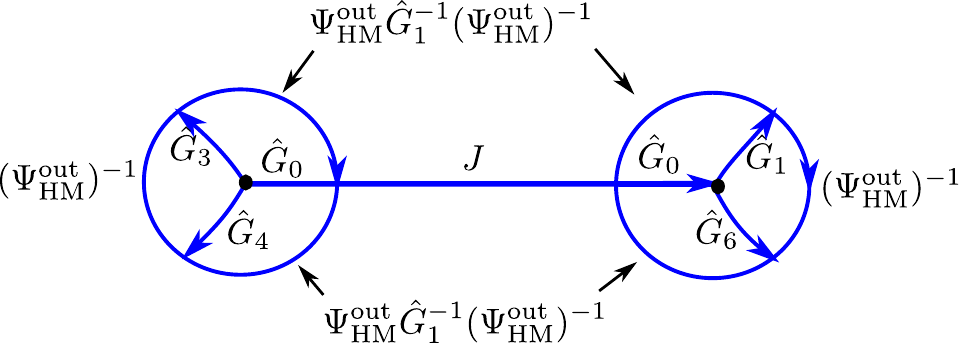}
\caption{The jump contours and jump matrices for the RH problem solved by $\hm$.  The radius for the two circles is $r$.}
\end{figure}

We define local parametrices near $\pm \alpha$:
\begin{align*}
\hma &= \begin{choices} I \when -\piot < \arg(z-a) < \piot\\
\hat G_1^{-1} \when \piot < \arg(z-a) < \pi\\
\hat G_1 \when - \pi < \arg(z-a) < - \piot
\end{choices},\\
\hmna &= \begin{choices} I \when \tpiot < \arg (z+a) < \pi \text{ or } - \pi < \arg(z+a) < - \tpiot\\
\hat G_1^{-1} \when 0 < \arg(z+a) < \tpiot\\
\hat G_1 \when -\tpiot < \arg(z+a) < \tpiot \end{choices}.
\end{align*}
We are ready to define the global parametrix.  Given $r > 0$ define
\begin{align*}
\hm = \begin{choices} \hma \when |z-a| < r\\
\hmna \when |z+a| < r\\
\hmout \when |z+a| > r \text{ and } |z-a| > r
\end{choices}.
\end{align*}
It follows that $\hm$ satisfies the RH problem shown in Figure~\ref{HM4}.

\begin{figure}[ht]\label{HM35}
\centering
\subfigure[]{\includegraphics[width=.45\linewidth]{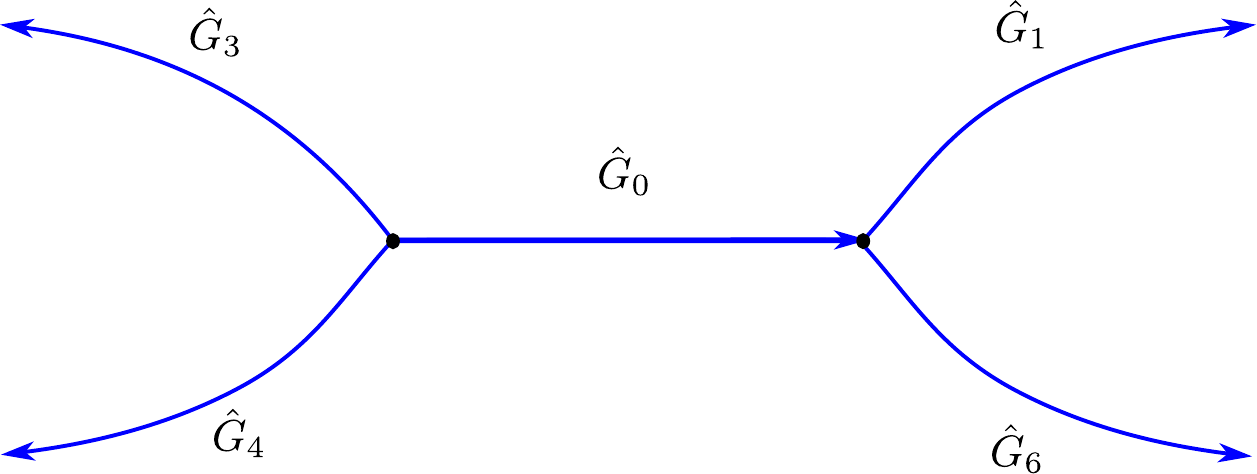}\label{HM3}}
\subfigure[]{\includegraphics[width=.45\linewidth]{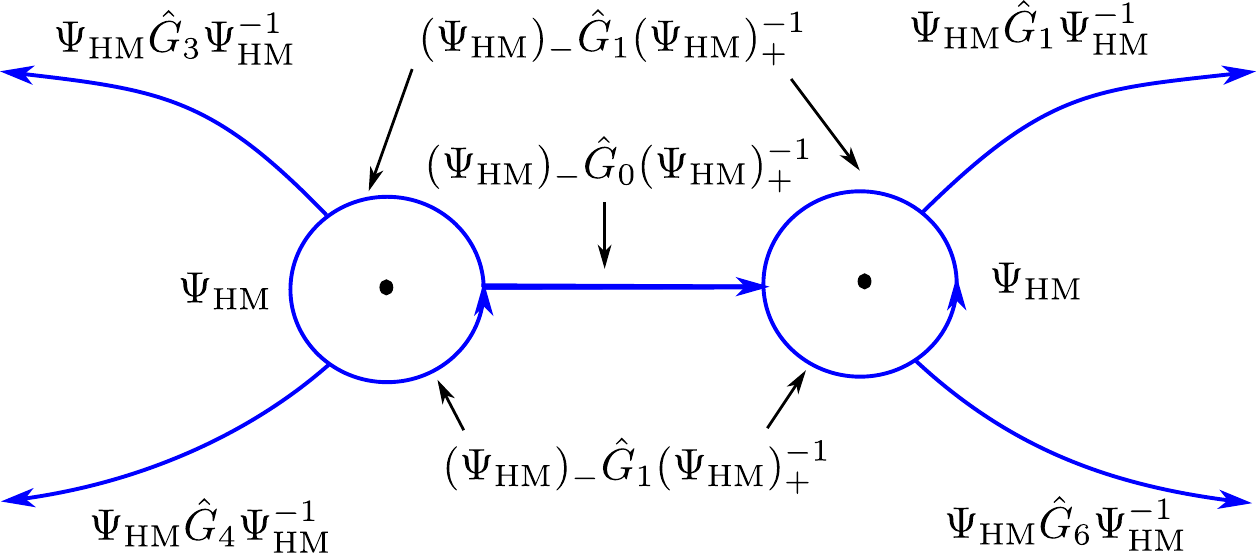}\label{HM5}}
\caption{The final deformation of the RH problem for negative $x$, with Assumption \ref{s1s3}.  The black dots represent $\pm \alpha$.  (a) After conjugation by $\Theta$. (b) Bounding the contours away from the singularities of $g$ and $\beta$ using $\hm$.}
\end{figure}

Let $\Phi$ be the solution of the RH problem shown in Figure~\ref{HM3}.  It follows that $\Delta = \Phi \hm^{-1}$ solves the RH problem shown in Figure~\ref{HM5}.
The RH problem for $\Delta$ has jump matrices that decay to the identity away from $\pm \alpha$.  We use Assumption~\ref{scalings} to determine that we should use $r = |x|^{-1}$.  We solve the RH problem for $\Delta$ numerically.  To compute the solution of \Painleve\ II we use the formula
\begin{align*}
P_{II}(\pm i,0,\mp i;x) = 2i \lim_{z\goto \infty} z \Delta(z)_{12}.
\end{align*}

\begin{figure}[ht]\label{HMPlotting}
\centering
\subfigure[]{\includegraphics[width=.8\linewidth]{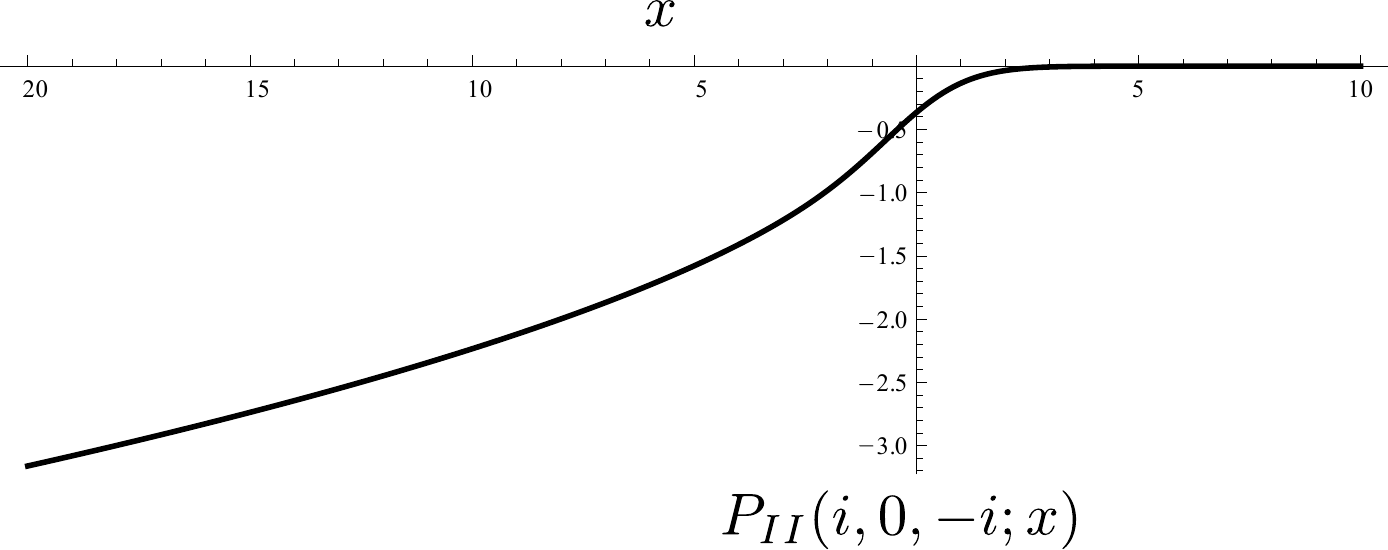}\label{HMFull}}\\
\subfigure[]{\includegraphics[width=.45\linewidth]{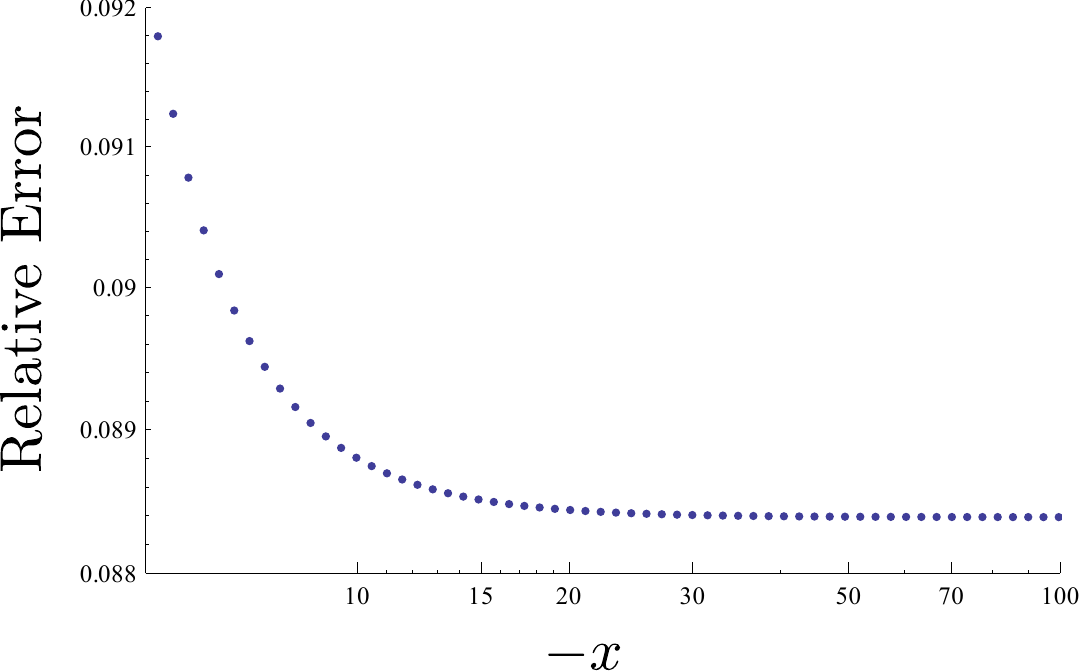}\label{HMRelError}}
\caption{Plotting and analysis of the numerical approximation of $P_{II}(i,0,-i;x)$.  (a) $P_{II}(i,0,-i;x)$ for positive and negative $x$.  For small $|x|$ we solve the undeformed RH problem. (b) A verification the numerical approximation using the asymptotics \eqref{PIIasymptotics}.}
\end{figure}

See Figure~\ref{HMFull} for a plot of the Hastings--McLeod solution with $s_1 = i$.  To verify our computations we may we use the asymptotics \cite{FokasPainleve}:
\begin{align}\label{PIIasymptotics}
P_{\rm II}(i,0,-i;x) \sim - \sqrt{\frac{-x}{2}} + \bigo\left(x^{-5/2}\right).
\end{align}
We define
\begin{align*}
\left|\frac{P_{\rm II}(i,0,-i;x)+\sqrt{\frac{-x}{2}}}{x^{-5/2}}\right|,
\end{align*}
to be the \emph{relative error} which should tend to a constant for $x$ large and negative.  We demonstrate this in Figure~\ref{HMRelError}.

\begin{remark} Since $\beta$ has unbounded singularities we expect that a similar issue as in \secref{unbounded} will arise. We do not go though the details of this but this approach produces accurate numerics for all $x$ on the real line.\end{remark}

\References

\ends